\documentclass[aps,prd,twocolumn,10pt]{revtex4-2}
\usepackage[utf8]{inputenc}
\usepackage{enumitem}
\usepackage{mathtools}
\usepackage{ragged2e}
\usepackage{booktabs,tabularx}
\usepackage{array}
\usepackage{entanglement_QED}
\usepackage[colorlinks]{hyperref}
\usepackage[capitalise]{cleveref}
\usepackage[caption=false]{subfig}
\usepackage{tikz}

\newenvironment{subcaptions}[1]{%
        \begin{tikzpicture}[every node/.style={inner sep=0}]%
        \node[use as bounding box,anchor=north west] (image) at (0,0) {#1};%
        \begin{scope}[x={(image.north east)},y={(image.south west)}]%
    }{%
        \end{scope}\end{tikzpicture}%
}
\newcommand*{\oversubcaption}[2]{\node[anchor=north west, transparent] at (#1) {\subfloat[#2]{\hspace{2ex}}};}
\newcommand\sref[1]{(\protect\subref{#1})}

\newcolumntype{C}{>{\;$}c<{$\;}}

\def\emdash{\,---\,}

\begin{document}

\title{Entanglement generation in \boldmath\texorpdfstring{$(1+1)D$}{(1+1)D} QED scattering processes}

\author{Marco Rigobello}
\email{marco.rigobello.2@phd.unipd.it}
\affiliation{Dipartimento di Fisica e Astronomia ``G. Galilei'', via Marzolo 8, I-35131, Padova, Italy}
\affiliation{Padua Quantum Technologies Research Center, Universit\`a degli Studi di Padova.}
\affiliation{INFN, Sezione di Padova, via Marzolo 8, I-35131, Padova, Italy}

\author{Simone Notarnicola}
\affiliation{Dipartimento di Fisica e Astronomia ``G. Galilei'', via Marzolo 8, I-35131, Padova, Italy}
\affiliation{Padua Quantum Technologies Research Center, Universit\`a degli Studi di Padova.}
\affiliation{INFN, Sezione di Padova, via Marzolo 8, I-35131, Padova, Italy}

\author{Giuseppe Magnifico}
\affiliation{Dipartimento di Fisica e Astronomia ``G. Galilei'', via Marzolo 8, I-35131, Padova, Italy}
\affiliation{Padua Quantum Technologies Research Center, Universit\`a degli Studi di Padova.}
\affiliation{INFN, Sezione di Padova, via Marzolo 8, I-35131, Padova, Italy}

\author{Simone Montangero}
\affiliation{Dipartimento di Fisica e Astronomia ``G. Galilei'', via Marzolo 8, I-35131, Padova, Italy}
\affiliation{Padua Quantum Technologies Research Center, Universit\`a degli Studi di Padova.}
\affiliation{INFN, Sezione di Padova, via Marzolo 8, I-35131, Padova, Italy}

\begin{abstract}
    We study real-time meson-meson scattering processes in $(1+1)$-dimensional QED by means of Tensor Networks.
    We prepare initial meson wave packets with given momentum and position introducing an approximation based on the free fermions model.
    Then, we compute the dynamics of two initially separated colliding mesons, observing a rich phenomenology as the interaction strength and the initial states are varied in the weak and intermediate coupling regimes.
    Finally, we consider elastic collisions and measure some scattering amplitudes as well as the entanglement generated by the process.
    Remarkably, we identify two different regimes for the asymptotic entanglement between the outgoing mesons: it is perturbatively small below a threshold coupling, past which its growth as a function of the coupling abruptly
    accelerates.
\end{abstract}

\maketitle

The investigation of fundamental interactions is one of the more challenging research fields in physics from a theoretical, experimental and computational point of view.
The Standard Model (SM) of particles physics describes the fundamental components of matter as quantum fields, whose interactions are set by the system invariance under specific gauge transformation groups \cite{Weinberg2004MakingStandardModel}.
Many unveiled phenomena, such as  the matter–antimatter asymmetry, the origin of dark matter and dark energy, as well as a complete understanding of the Quantum Chromodynamics (QCD) phase diagram, motivate intense research for physics in and beyond the SM \cite{Lykken2010StandardModel}.
In particle accelerators, scattering events are extensively exploited to investigate the properties of fundamental particles and their interactions.
\begin{figure}[t!]
    \centering
    \includegraphics[width=0.48\textwidth]{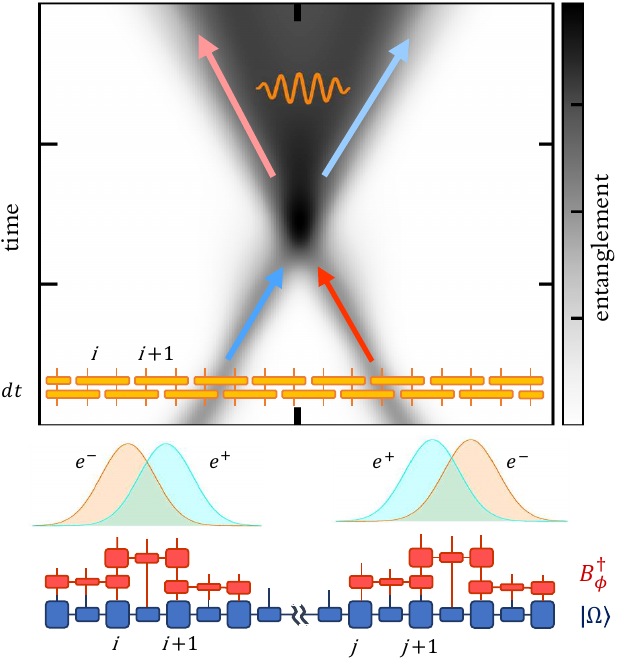}
    \caption{\label{fig:intro}%
        The ground state of the Schwinger model is computed and expressed as an MPS (blue bottom squares).
        Then a set of MPOs (red squares) is applied  to prepare a gauge invariant initial state made up of two spatially separated meson wave packets with opposite electric field orientations.
        Indexes $i$ and $j$ indicate lattice sites on which matter and antimatter field components are defined, while the electric field is defined on the links between sites.
        The two mesons evolve under the Hamiltonian of Eq. (\ref{eq:Hqed}) with TEBD (pictorially represented by yellow boxes) and their scattering is studied.
        Note the persistent entanglement between them after the scattering process. Numerical parameters: $\spc*=1$, $\mass=1.0$, $\cpl=0.14$.
    }
\end{figure}
In this scenario, numerical simulations play a fundamental role to compare experimental results and theoretical predictions.
The most common approach for the numerical simulation of quantum gauge theories relies on Lattice Gauge Theory (LGT) models \cite{Rothe2012LatticeGaugeTheories,Montvay1997QuantumFieldsLattice}, in which quantum fields are defined on a discrete space-time lattice. Within this framework, Monte Carlo techniques are the established tool to predict scattering amplitudes \cite{Buckley2019MonteCarloEvent}.
Even though Monte Carlo simulations of LGTs have been able to correctly interpret an impressive number of experimental observations, they suffer from the notorious sign problem, that severely limits the study of the finite-density region of the QCD phase diagram and the real-time dynamics \cite{Dowling2007MonteCarloTechniques,Gattringer2016ApproachesSignProblem,Fukushima2010PhaseDiagramDense,Endroedi2011QcdPhaseDiagram}.
In the last decades Tensor Network (TN) methods \cite{Schollwoeck2011DensityMatrixRenormalization,Orus2014PracticalIntroductionTensor, Biamonte2017TensorNetworksNutshell,Biamonte2019LecturesQuantumTensor,Montangero2018IntroductionTensorNetwork,Silvi2019TensorNetworksAnthology}, thanks to their capability to efficiently compress the information contained in quantum many-body states \cite{Kluemper1993MatrixProductGround,Verstraete2004RenormalizationAlgorithmsQuantum,  Verstraete2006CriticalityAreaLaw, Vidal2007EntanglementRenormalization, Verstraete2008MatrixProductStates, Evenbly2009EntanglementRenormalizationTwo, Gerster2014UnconstrainedTreeTensor, Shi2006ClassicalSimulationQuantum, Tepaske2021ThreeDimensionalIsometric, Vlaar2021SimulationThreeDimensional, Felser2021EfficientTensorNetwork}, have emerged as a sign-problem free tool to perform numerical simulations of fermionic LGT models \cite{Rico2014TensorNetworksLattice,Silvi2014LatticeGaugeTensor,Tagliacozzo2014TensorNetworksLattice}.
TNs have been successfully applied to study $(1+1)$-dimensional LGTs, characterising the phase diagram and the dynamics in Abelian \cite{ Byrnes2002DensityMatrixRenormalization, Pichler2016RealTimeDynamics, Buyens2017RealTimeSimulation, Banuls2017DensityInducedPhase, Magnifico2019SymmetryProtectedTopological, Magnifico2019ZnGaugeTheories, Funcke2020TopologicalVacuumStructure, Zache2021AchievingContinuumLimit} and non-Abelian \cite{Silvi2019TensorNetworkSimulation, Silvi2017FiniteDensityPhase, Banuls2017EfficientBasisFormulation} models.
Recently, TNs have also used to study the phase diagram of $(2+1)$- and $(3+1)$-dimensional Abelian LGTs in different regimes of model parameters \cite{Bender2020RealTimeDynamics, Zohar2015FermionicProjectedEntangled, Huang2019DynamicalQuantumPhase, Emonts2020VariationalMonteCarlo, Zohar2021WilsonLoopsArea, Nyhegn2021ZnLatticeGauge} and also in the finite-density scenario \cite{Felser2020TwoDimensionalQuantum,Magnifico2020LatticeQuantumElectrodynamics}.

Here, we perform TN simulations by using Matrix Product States (MPS) to investigate the real-time dynamics of scattering events in a $(1+1)$-dimensional Abelian LGT in the Hamiltonian formulation \cite{Kogut1975HamiltonianFormulationWilsons,Kogut2003PhasesQuantumChromodynamics}.
In particular, we  set up a protocol for the initialization of wave packets of asymptotic isolated states and observe their scattering dynamics.
We focus on the Schwinger model, i.e. \qed{}, which describes electrons and positrons interacting on a 1D lattice via a scalar electric field \cite{Kogut1975HamiltonianFormulationWilsons}.
Since the Schwinger model does not admit stable isolated-charge excitations \cite{Nakanishi1978AsymptoticCompletenessConfinement}, we compute the scattering between two meson wave packets, each of them composed by an electron and a positron, sufficiently separated in space to be initially uncorrelated, as pictorially shown in Fig. \ref{fig:intro}.
In order to prepare the initial state for the dynamics, we compute the ground state of the model Hamiltonian  by using the Density Matrix Renormalization Group (DMRG)  algorithm  with MPS representation \cite{White1992DensityMatrixFormulation,White1993DensityMatrixAlgorithms,Schollwoeck2005DensityMatrixRenormalization,Schollwoeck2011DensityMatrixRenormalization}. Then, we apply a set of Matrix Product Operators (MPO)  which create the mesons \cite{Pirvu2010MatrixProductOperator}.
This approach allows us to fix the space and momentum center-of-mass position for each meson independently, as well as their widths.
After the preparation of the initial state, the dynamics is computed via the Time Evolving Block Decimation (TEBD) algorithm \cite{Vidal2003EfficientClassicalSimulation,Vidal2004EfficientSimulationOne}.
We analyze the scattering products in terms of the final momenta, by computing the structure factor of the final state and some scattering amplitudes, in order to better characterize the final state and compare it to the initial one.
Finally, we focus on the entanglement production during the process:
the initial entanglement between the two mesons is zero before the scattering event, as expected.
After the scattering, we observe two very well-separated behaviors as the Hamiltonian parameters are varied:
one in which the entanglement at long times is perturbatively small with respect to the electric field coupling, and another in which a significant amount of entanglement persists asymptotically in time.

Our work paves the way to future insights into the role played by entanglement in scattering processes in LGTs, extending previous analytical studies and numerical results obtained for spin models \cite{Peschanski2016EntanglementEntropyScattering,Casini2009EntanglementEntropyFree, Vanderstraeten2014SMatrixMatrix,Surace2021ScatteringMesonsQuantum,Karpov2020SpatiotemporalDynamicsParticle,Magoni2021EmergentBlochOscillations}. Moreover, TNs provide the ideal framework and language to make a link between our protocol and experimental quantum simulations and computations of LGTs. Indeed, motivated by the improvements in experimental techniques for the manipulation of isolated quantum many-body systems \cite{Bruzewicz2019TrappedIonQuantum,Scholl2020ProgrammableQuantumSimulation,Omran2019GenerationManipulationSchrodinger,Browaeys2020ManyBodyPhysics,Ebadi2020QuantumPhasesMatter,Wintersperger2020RealizationAnomalousFloquet,Bluvstein2021ControllingQuantumMany,Preskill2018QuantumComputingNisq}, a plethora of analog and digital quantum simulators proposals has come out to implement Abelian and non-Abelian LGTs into experimental setups \cite{Banuls2020SimulatingLatticeGauge,Celi2020EmergingTwoDimensional,Jordan2012QuantumAlgorithmsQuantum,Jordan2014QuantumAlgorithmsFermionic,Jordan2018BqpCompletenessScattering}.
In this framework, our TNs protocol could also serve as a benchmarking toolbox for realistic scattering implementations on quantum hardware.

The manuscript is organized as follows.
In \cref{sec:models} we illustrate the models involved in our simulations\emdash{}namely, free lattice fermions and \qed{}.
\Cref{sec:methods} contains an overview of the generic tools employed in our investigations.
In \cref{sec:initial_state} we present the protocol that we use to model the initial state of a scattering experiment and we introduce operators that prepare wave packets of fermion and antifermions (free theory) and mesons (\qed{}).
In \cref{sec:dynamics,sec:entanglement} we analyze the simulations of some meson-meson collisions.
\Cref{sec:dynamics} focuses on the collision phenomenology and includes the computation of some scattering amplitudes.
\Cref{sec:entanglement} focuses on the entanglement content of the system during the scattering process and includes an estimation of the entanglement between the final products that is generated by the interactions.

\togglefalse{latunits}

\section{Models}\label{sec:models}
We adopt units where $\hbar=c=1$ and consider a discretization of one-dimensional space as a uniform chain $\latx$ of an even number $\cells\in2\integers$ of sites separated by lattice spacing $\spc$,
\begin{math}
    \latx = \spc* \left\lbrace 0, \ldots, \cells-1 \right\rbrace
\end{math}. Lattice positions are denoted by
\begin{math}
    \xx[1],\xx[2],\xx[3]\in\latx
\end{math};
lattice momenta are
\begin{math}
    \kk[1],\kk[2],\kk[3]\in\latk
\end{math},
\begin{equation}\label{eq:reciprocal_lattice}
    \latk=
    \kspc
    \left\{
    -\frac{\cells}{2},
    -\frac{\cells}{2}+1,
    \ldots,
    \frac{\cells}{2}-1
    \right\}
    \cong\kspc\Zn[\cells]
    \,.
\end{equation}
We now give a possible lattice definition of the $(1+1)$-dimensional theories of free relativistic Dirac fermions and of quantum electrodynamics; the latter being the main subject of this work.

\subsection{Free fermions}
The Kogut-Susskind discretization of a mass $\mass$ free relativistic Dirac field in $(1+1)$-spacetime dimensions is described by the Hamiltonian \cite{Kogut1975HamiltonianFormulationWilsons,Banks1976StrongCouplingCalculations,Susskind1977LatticeFermions}
\begin{equation}\label{eq:Hfree_x}
    H = \sumx \bigg[
        \frac{i}{2\spc*} \stgx[\xx+\spc]* \stgx + \hc + \mass \stgsgn \stgx*\stgx \bigg]
    \,.
\end{equation}
The staggered fermion $\stgx$ degrees of freedom on even and odd lattice sites correspond respectively to the upper and lower components of the $(1+1)$-dimensional Dirac spinor field;
hence the alternating sign in front of the mass term in \cref{eq:Hfree_x} \cite{Susskind1977LatticeFermions}.
The fields $\stgx$ and $\stgx*$ satisfy canonical anticommutation relations \cite{Banks1976StrongCouplingCalculations}:
\begin{equation}\label{eq:free_commutators}
    \acomm*{\stgx[2]}{\stgx[3]*} = \iftoggle{latunits}{}{\spc*^{-1}} \delta\indices{_{\xx[2]\xx[3]}}
    \,,
\end{equation}
while other fundamental anticommutators vanish.
The Jordan-Wigner transformation \cite{Jordan1928UeberDasPaulische,Susskind1977LatticeFermions} provides an irreducible matrix representation of this algebra in the local occupation number $\occ=\spc*\,\stgx*\stgx$ eigenbasis $\ket*{\freebasis}$, $\freebasis \in \{0,1\}^{\cells}$.

As its continuum counterpart, the Hamiltonian in \cref{eq:Hfree_x} has a global $\Uone$ symmetry, generated by the particle number conserved charge $Q=\sumx\stgx*\stgx$ \cite{Susskind1977LatticeFermions}.
In the thermodynamic limit or for periodic boundaries the theory is also translation invariant.
Valid translations, however, are generated by shifts by two lattice steps \cite{Susskind1977LatticeFermions}.
As a consequence, the effective momentum space of staggered fermions is
\begin{math}
    \stglatk = \latk \cap [-\pi/2\spc*, \pi/2\spc*[
\end{math};
we denote momentum sums restricted to this sublattice by $\stgsumk*[]$.
Moreover, it is sometimes convenient to isolate the even and odd sublattices, namely
\begin{math}
    \sys{E}=\latx\cap a(2\integers)
\end{math}
and
\begin{math}
    \sys{O}=\latx\cap a(2\integers+1)
\end{math}.

\subsection{QED}
A discretization of $(1+1)$-dimensional Quantum Electrodynamics (\qed, also known as massive Schwinger model \cite{Schwinger1962GaugeInvarianceMass.,Lowenstein1971QuantumElectrodynamicsTwo}), is obtained promoting the global $\Uone$ symmetry of \cref{eq:Hfree_x} to a gauge symmetry.
In the Kogut-Susskind formalism, the theory is defined by the Hamiltonian \cite{Kogut1975HamiltonianFormulationWilsons,Banks1976StrongCouplingCalculations,Susskind1977LatticeFermions}
\begin{multline}\label{eq:Hqed}
    H = \sumx \bigg[
    \frac{i}{2\spc*} \stgx[\xx+\spc]*\cmpR \stgx + \hc \\
    {} + \mass \stgsgn \stgx*\stgx \\
    {} + \frac{\cpl^2}{2} (\eleR)^2
    \bigg]
    \,.
\end{multline}
\Cref{eq:Hqed} involves matter degrees of freedom, $\stgx$ and $\stgx*$, living on the lattice sites;
as well as unitary gauge parallel transporters $\cmpR$
and Hermitian electric field $\cpl\eleR$ operators,
acting on local Hilbert spaces associated to the links between neighboring sites \cite{Banks1976StrongCouplingCalculations}.
Both the bare mass $\mass$ and coupling $\cpl$ parameters have the dimension of a mass, thus the ratio $\cpl/\mass$ is adimensional and can be used to quantify the strength of the interaction and interpolate between the weak ($\cpl\ll\mass$) and strong ($\cpl\gg\mass$) coupling limits \cite{Coleman1976MoreMassiveSchwinger}.
In addition to the matter anticommutator from \cref{eq:free_commutators}, we have the non-vanishing fundamental commutator \cite{Banks1976StrongCouplingCalculations}
\begin{equation}\label{eq:commutators}
    \comm*{\cmpR[2]}{\eleR[3]} = \delta\indices{_{\xx[2]\xx[3]}}\cmpR[2]
    \,,
\end{equation}
showing that parallel transporters act as lowering operators for the electric field on the same lattice link.
An irreducible representation of \cref{eq:commutators} on a given link (left implicit) reads \cite{Banks1976StrongCouplingCalculations,Konishi2009QuantumMechanicsNew}
\begin{equation}\label{eq:link_irrep}
    \eleSym\ket{\eleEig} = \eleEig\ket{\eleEig}
    , \ \
    \cmpSym\ket{\eleEig} = \ket{\eleEig-1}
    , \ \
    \eleEig \in \spectrum(\eleSym) =  \integers
    \, ,
\end{equation}
where $\spectrum(\eleSym)$ denotes the spectrum of the operator $\eleSym$.
There exist infinite other unitarily inequivalent representations related to \cref{eq:link_irrep} by a shift of the electric field:
$\eleSym\to\eleSym+\delta\eleEig$, $0 < \delta\eleEig < 1$ \cite{Konishi2009QuantumMechanicsNew}.
Here we assume $0\in\spectrum(\eleSym)$ on all links.
The representation in \cref{eq:link_irrep} extends trivially to the whole chain.
Time independent $\Uone$ gauge transformations are implemented via $\exp(\gauge)$, with $\param\in\reals^{\cells}$ parametrizing the local transformation generated by the Gauss operator $\gauss$ \cite{Kuehn2014QuantumSimulationSchwinger,Henneaux1992QuantizationGaugeSystems},
\begin{equation}\label{eq:gauss}
    \gauss = \spcfrac{\eleR - \eleL} - \stgx*\stgx + \frac{1-\stgsgn}{2\spc*}
    \,,
\end{equation}
which commutes with the Hamiltonian.
As a boundary condition, we identify the electric field at the left and at the right of the chain with the null operator, thus restricting our analysis to the sector with total charge zero.
Physical states
\begin{math}
    \ket{\pphys}\in\hilphys\subset\hil
\end{math}
are required to be gauge invariant \cite{Kogut1975HamiltonianFormulationWilsons,Henneaux1992QuantizationGaugeSystems}, namely to satisfy
\begin{equation}\label{eq:U1_phys}
    \exp\relax\Big(\gauge\Big)\ket{\pphys} = \ket{\pphys}
    \quad
    \forall\;\param
    \,.
\end{equation}
The physical state condition is equivalent to Gauss law, i.e.,
\begin{math}
    \spc*\gauss\ket{\pphys} = 0
\end{math}
for all $\xx$.

In $1+1$ dimensions and with $\eleboundary$, given an arbitrary configuration $\freebasis$ of the matter fields there is one and only one configuration $\elebasis$ of the link degrees of freedom complying with Gauss law; precisely
\begin{equation}
    \eleR{\eigStyle}
    = \sumx*[\xx[2]<\xx]\left[ \occ{\eigStyle}[2] - \frac{1-\stgsgn}{2} \right]
    \,.
\end{equation}
It follows that $\ket*{\freebasis}$ provides a basis for $\hilphys$, which is thus unitarily equivalent to the Hilbert space of free staggered fermions $\hilfree$.
Consequently, a free theory operator $O$ also defines an operator on $\hilphys$ that can be extended to a (gauge invariant) dressed operator $\bar{O}$ on the whole $\hil$.
Conversely, as far as gauge independent properties of the model are concerned, gauge invariant \qed{} operators can be expressed in the $\ket*{\freebasis}$ basis.
For instance, in this basis the Hamiltonian becomes
\begin{equation}\label{eq:Hqed_integrated}
    H =
    H_{\text{\free{}}}
    + \frac{\spc*\cpl^2}{2}\sumx*\bigg[
        \sumx*[2]^{\xx} \bigg(\occ[2] - \frac{1-\stgsgn[2]}{2}
        \bigg)
        \bigg]^{2}
    \,,
\end{equation}
where $H_{\text{\free{}}}$ is the free staggered fermion Hamiltonian from \cref{eq:Hfree_x} \cite{Hamer1997SeriesExpansionsMassive}.
With this procedure the link degrees of freedom have been removed\emdash{}namely, integrated out\emdash{}and the gauge redundancy of the model has been eliminated.

The reformulation of the theory in \cref{eq:Hqed_integrated} is convenient for some derivations carried out in this work.
In numerical simulations, instead, we keep the (redundant) link degrees of freedom to avoid long-range interactions.
Their infinite dimensional local Hilbert spaces are truncated introducing a cutoff $\eleMax\in\naturals$ in the $\abs{\eleSym}$ spectrum and imposing
\begin{math}
    \cmpSym\ket{-\eleMax} = \ket{+\eleMax}
\end{math}
on each link; in this way
$\spectrum(\eleSym)$
is identified with $\Zn[\eleLvl]$, $\eleLvl=2\eleMax+1$.
This truncation spoils the commutator in \cref{eq:commutators}, explicitly breaking the $\Uone$ gauge invariance of the model down to the $\Zn[\eleLvl]$ residual symmetry group of transformations with parameters
\begin{math}
    \left.\param\in ({2\pi}/{\eleLvl}) \integers\right.
\end{math}
\cite{Horn1979HamiltonianApproachZn,Elitzur1979PhaseStructureDiscrete,Kuehn2014QuantumSimulationSchwinger,Notarnicola2015DiscreteAbelianGauge, Ercolessi2018PhaseTransitionsZn}.
The physical state condition in \cref{eq:U1_phys} is accordingly weakened and the Gauss law only holds modulo $\eleLvl$.
Namely, the physical Hilbert subspace $\hilphys$ is spanned by the occupation number and electric field eigenstates satisfying
\begin{math}\label{eq:ZN_phys}
    \spc*\gauss[2]\ket*{\qedbasis} \in (\eleLvl\integers)\ket*{\qedbasis}
\end{math}
for all $\xx[2]$.
Our simulations rely on an $\eleLvl=7$ truncation.
It has been shown that $\eleLvl=3$ gives already an excellent approximation of the exact ground state of \qed{} \cite{Kuehn2014QuantumSimulationSchwinger,Buyens2014MatrixProductStates,Buyens2017FiniteRepresentationApproximation};
even though the $\Zn[3]$ model is also capable of reproducing accurately some dynamical processes of the untruncated theory, the quality of the approximation depends on the value of the model parameters and the specific process under consideration \cite{Notarnicola2020RealTimeDynamics,Magnifico2020RealTimeDynamics}.
Monitoring the system state $\ket{\P}$ during $\Zn[\eleLvl]$ ($\eleLvl=3,5,7$) scattering simulations, we find that the fraction $\chi_{\text{trunc}}$ of configurations affected by truncations of the electric field is below the numerical precision of the simulation for $\eleLvl = 7$.
Precisely, $\chi_{\text{trunc}} = 1 - \ev{\Pi_{\Uone}}{\P} \le \order*{10^{-10}}$, where $\Pi_{\Uone}$ is the projector on the Hilbert subspace of states complying with the $\Uone$ Gauss law.

\section{Methods}\label{sec:methods}

In this work, Tensor Network (TN) methods \cite{Biamonte2017TensorNetworksNutshell,Biamonte2019LecturesQuantumTensor,Montangero2018IntroductionTensorNetwork,Silvi2019TensorNetworksAnthology,Hauschild2018EfficientNumericalSimulations} are employed to tackle the exponential growth of the many body Hilbert space of the lattice theories introduced in \cref{sec:models}.
Specifically, we use the Matrix Product State (MPS) \cite{Rommer1997ClassAnsatzWave} and Matrix Product Operator (MPO) \cite{Pirvu2010MatrixProductOperator} ansätze to represent states and operators.
In conjunction, we use the MPS implementation of (i) the (two-site) Density Matrix Renormalization Group (DMRG) algorithm to perform variational optimizations; and (ii) the (fourth-order) Time Evolving Block Decimation (TEBD) algorithm to compute Trotterized time evolutions \cite{White1992DensityMatrixFormulation,White1993DensityMatrixAlgorithms,Schollwoeck2005DensityMatrixRenormalization,Schollwoeck2011DensityMatrixRenormalization,Vidal2004EfficientSimulationOne,Verstraete2004MatrixProductDensity}.
In order to recast the \qed{} Hamiltonian of \cref{eq:Hqed} in a nearest-neighbor Hamiltonian, as required by TEBD, we fuse the Hilbert spaces associated to a site $\hil_{\xx}$ and to the subsequent link $\hil_{\xx,\xx+\spc}$ in a single (larger) local computational Hilbert space $\hil_{\xx}\otimes\hil_{\xx,\xx+\spc}\to\hil_{\xx}$.

Finally, we exploit some of the symmetries of the problem to improve the efficiency and the numerical precision of our simulations.
Using a symmetric MPS ansatz \cite{Montangero2018IntroductionTensorNetwork,Silvi2014LatticeGaugeTensor,Silvi2019TensorNetworksAnthology}, we constrain the ground state search and the dynamics in the desired charge sector.
Specifically, we impose\emdash{}exactly\emdash{}the conservation of the $\Uone$ global charge $Q$ and, for the \qed{} case, also the vanishing of the total $\Zn[\eleLvl]$ Gauss charge on even and odd sublattices
, i.e.,
$\sumx[\xx\in\sys{E}]\gauss$ and
$\sumx[\xx\in\sys{O}]\gauss$ \cite{Tschirsich2019PhaseDiagramConformal,RishonsNote}.

Schematically, the presented results are obtained as follows (as depicted in \cref{fig:intro}):
\begin{enumerate}
    \item Determine an MPS representation of the ground state $\ket{\vac}$ of the model via DMRG.
    \item Prepare an initial particle wave packets MPS $\ket{\P}$ acting on $\ket{\vac}$ with MPOs, as discussed in \cref{sec:initial_state}.
    \item Determine the time evolution $\ket{\P(t)} = e^{-iHt}\ket{\P}$, via the TEBD algorithm.
\end{enumerate}
During the evolution we monitor the expectation values $\ev{O}_t=\ev{O}{\P(t)}$ of relevant observables $O$,
such as the energy $\ev*{H}_t=\ev*{H}$ and the charge density $\ev*{\stgx*\stgx}_t$ or, equivalently, the mass energy density $\hmass$,
\begin{equation}
    \hmass_{\xx}(t) = \mass\stgsgn\ev*{\stgx*\stgx}_t
    \,.
\end{equation}
For \qed{} we also measure the electric field $\cpl\ev*{\eleR}_t$ and its energy density $\helec$,
\begin{equation}
    \helec_{\xx}(t) = \frac{\cpl^2}{2}\ev*{(\eleR)^2}_t
    \,.
\end{equation}
Finally, we characterize the entanglement content of the system by computing the Von Neumann entanglement entropy $\ent$ associated to every bipartition of the chain in two subsystems,
\begin{math}
    \sys{L} = \{ \xx[2]<\xx \}
\end{math}
and
\begin{math}
    \sys{R} = \{ \xx[2]\ge\xx \}
\end{math};
namely,
\begin{equation}
    \ent(\xx,t) = -\tr\big[\densmat(t)\log_2\densmat(t)\big]
    \,,
\end{equation}
where $\densmat(t)$ is the reduced density matrix of one subsystem, e.g., the partial trace of $\op{\P(t)}$ over
\begin{math}
    \bigotimes_{\xx[2]\in\sys{L}} \hil_{\xx[2]} \vspace{2pt}
\end{math}.

Unless otherwise stated, we report the deviation from the vacuum (ground state) expectation value of all the above quantities.
Moreover, we use the lattice spacing to fix the overall length scale, i.e., we set $\spc=1$, and all simulations are carried out with open boundary conditions.

Numerical errors have been kept under control.
The MPS compression is achieved truncating singular values below $10^{-6}$, resulting in an overall maximum MPS bond dimension of $991$;
the Trotterization of the evolution (we exploit a fourth-order Suzuki-Trotter decomposition \cite{Suzuki1991GeneralTheoryFractal} with time step $\delta t = 0.05$) is responsible for an overall $\order*{10^{-3}}$ error on the final state; while
the truncation of the electric field spectrum ($\abs{\cpl\eleSym} \le 7\cpl$) is discussed in \cref{sec:models}.
The convergence of some relevant quantities\emdash{}e.g., of the midchain entanglement entropy\emdash{}with the MPS bond dimension has been verified.

\section{Initial State Preparation}\label{sec:initial_state}

The state of a system undergoing a scattering process long before the collision is populated with wave packets of stable particle excitations localized in far distant space regions \cite{Haag1996LocalQuantumPhysics,Weinberg1995QuantumTheoryFields}.
A prerequisite for the description of a scattering experiment is thus the identification of the model's particles and their asymptotic-times dynamics \cite{Hannesdottir2020SMatrixMassless,Strocchi2013IntroductionNonPerturbative}.
In the absence of bound states and long range interactions, the aforementioned task consists in the computation of the plane wave solutions of the free theory associated to the degrees of freedom of the model \cite{Strocchi2013IntroductionNonPerturbative}.
The previous assumptions are trivially verified by a free theory, e.g. free fermions, but do not hold for \qed{} \cite{Nakanishi1978AsymptoticCompletenessConfinement,Coleman1975ChargeShieldingQuark,Lowenstein1971QuantumElectrodynamicsTwo,Abdalla2001NonPerturbativeMethods}.

In this Section we present and motivate our protocol for preparing the initial state of a \qed{} scattering simulation, providing an expression for the wave packet amplitudes and operators involved.
An MPO representation of these operators is given in \cref{app:MPOs}.
We consider initial states of uncorrelated particles\emdash{}i.e., particle wave packets described by independent amplitudes.
We start by illustrating how localized particle excitations are prepared in the lattice theory of free staggered fermions, as it provides useful insights into the \qed{} case.

\subsection{Free fermions}
In \cref{app:free_sol} we solve exactly the theory of free staggered fermions, presenting the Fock space structure of the Hilbert space.
Periodic lattice boundary conditions are assumed in the derivation in order to neglect open boundary effects.
Energy-momentum eigenstates with definite particle and antiparticle number are obtained acting with $\posk*$ and $\negk*$ operators on the vacuum $\ket{\freevac}$, i.e., the ground state of the model.
In \cref{eq:ab_def} we express the anticommuting creation operators $\posk*$ and $\negk*$ in terms of $\stgx*$ and $\stgx$.
However, these states are completely delocalized in space (and time) while realistic particle states are characterized by some space localization and are prepared acting with wave packets of $\posk*$ and $\negk*$, defined as
\begin{subequations}\label{eq:free_wavepacket_creation}
    \begin{align}
        \poswp* & = \stgsumk\wpk\posk* = \sumx\wpx^{\poswpSym}\stgx*
        \label{eq:fermion_wavepacket_creation}
        \,,                                                          \\
        \negwp* & = \stgsumk\wpk\negk* = \sumx\wpx^{\negwpSym}\stgx
        \label{eq:antifermion_wavepacket_creation}
        \,,
    \end{align}
\end{subequations}
for a fermion and an antifermion respectively, with momentum space probability density $\sabs*{\wpk}$,
\begin{equation}\label{eq:wavepacket_normalization}
    \stgsumk \sabs*{\wpk} = 1
    \,.
\end{equation}
Combined with the orthonormality condition in \cref{eq:fock_orthonormal}, \cref{eq:wavepacket_normalization} enforces the normalization of the prepared state.
The functions $\wpx^{\poswpSym,\negwpSym}$ in \cref{eq:free_wavepacket_creation} are specified in terms of $\wpk$ through \cref{eq:ab_def}.
The typical momentum space amplitude $\wpk$ is that of a Gaussian wave packet, namely
\begin{equation}\label{eq:gaussian_wavepacket}
    \wpk = \normalization_{\wpk[]} \:
    e^{-ik\xmean} \:
    e^{-(\kk - \kmean)^2 / 4\ksdev^2}
    \,;
\end{equation}
where the phase centers the wave packet in $\xmean$ in position space and the normalization constant $\normalization_{\wpk[]}$ is fixed by \cref{eq:wavepacket_normalization}.
In the continuum ($\ksdev\gg\spc*$) and thermodynamic limits, the $\sabs*{\wpk}$ given in \cref{eq:gaussian_wavepacket} approaches the probability density of a Gaussian momentum space distribution with mean $\kmean$ and standard deviation $\ksdev$ \cite{Szablowski2001DiscreteNormalDistribution}.

\begin{figure}
    \includegraphics{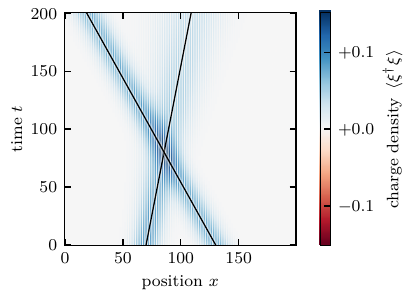}
    \caption{\label{fig:free}%
        Charge density during the free propagation of two different Gaussian fermion wave packets.
        The black lines in overlay are the trajectories of the wave packet peak predicted by the group velocity $\pulse'$.
    }
\end{figure}
The time evolution of a state consisting of two Gaussian fermions wave packets with different $\xmean$, $\kmean$ and $\ksdev$ is depicted in \cref{fig:free}.
The propagation speed of the two wave packets matches the expected result from the lattice group velocity, $\pulse'$, given by \cref{eq:dispersion}.
As \cref{fig:free} shows, the free theory phenomenology consists only of pure kinematics; in order to observe a nontrivial dynamics an interacting theory, such as \qed{}, has to be investigated.

\subsection{QED}
Continuum \qed{} has no free asymptotic charged states \cite{Coleman1975ChargeShieldingQuark,Lowenstein1971QuantumElectrodynamicsTwo,Abdalla2001NonPerturbativeMethods}: due to Gauss law and the linear rise of the Coulomb potential, typical of $1+1$ dimensions, the model exhibits a confining force and isolated charges correspond to states of infinite energy \cite{Abdalla2001NonPerturbativeMethods}.
The stable \qed{} particle states are thus neutral mesons, i.e., fermion-antifermion bound states.
In general, a creation operator for a lattice meson of momentum $\kk\in\latk$ reads
\begin{equation}\label{eq:meson_creation}
    \mesk* = \stgsumk[2;3] \kronk[(\kk[2]+\kk[3])\kk]* \, \meswfexact \, \posk{\dress}[2]* \negk{\dress}[3]*
    \,,
\end{equation}
for some function $\meswfexact$.
Here $\posk{\dress}*$ and $\negk{\dress}*$ are gauge invariant fermion and antifermion creation operators.
They are obtained (i) writing $\posk*$ and $\negk*$ in terms of the position space fields $\stgx*$ and $\stgx$ and (ii) dressing the latter with strings of unitary electric field rising or lowering operators acting on their right:
\begin{equation}\label{eq:dressing}
    \stgx* \to \stgx* \prod_{\xx[2]\geq\xx} \cmpR[2]*
    \,,\quad
    \stgx  \to \stgx  \prod_{\xx[2]\geq\xx} \cmpR[2]
    \,,
\end{equation}
to comply with Gauss law.
Unless appropriate conditions are imposed on $\meswfexact$, the operator $\mesk*$ in \cref{eq:meson_creation} creates excitations of definite momentum $\kk$ but not of definite energy.
Rather than seeking an approximate solution for \qed{} of the (notoriously difficult \cite{Broido1969GreenFunctionsParticle,Weinberg1995QuantumTheoryFields}) bound state problem, in this work we fix the functional form of $\meswfexact$ with an ansatz, namely
\begin{equation}\label{eq:mesonwf}
    \meswfexact
    = \meswf
    = \normalization_{\eta} \:
    e^{-{(\kk[3]-\kk[2])^2}/{4\dksdev^2}} \:
    e^{-i(\kk[3]-\kk[2]){\dxmean}/{2}}
    \,,
\end{equation}
where $\normalization_{\eta}$ is fixed requiring $\stgsumk\sabs*{\meswf[\kk]}=1$.
In \cref{eq:mesonwf} we switched to center of mass and relative coordinates of the fermion and antifermion constituents of the meson and assume $\meswfexact$ depends only on the latter.
Moreover, we require: (i) that the fermion and antifermion are located in close real space positions, $\xx[2]$ and $\xx[3]$, with average separation $\ev{\xx[3]-\xx[2]}=\dxmean$; (ii) that $(\kk[3]-\kk[2])$ follows a Gaussian probability distribution centered in $\dkmean=0$ and with standard deviation $\dksdev$.
Using \cref{eq:mesonwf} in place of the exact $\smash{\meswfexact}$ is equivalent to introducing some excitation in the bound state created by $\mesk*$.
As a consequence, some internal dynamics is to be expected.
We monitor this approximation a posteriori and set the simulation timescale shorter than the lifetime of the mesons.

\begin{figure}
    \begin{subcaptions}{\includegraphics{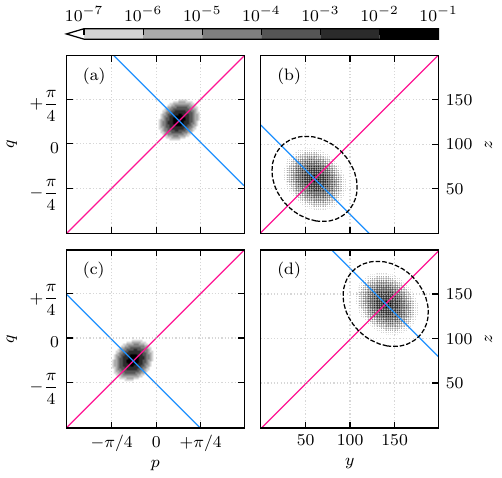}}
        \oversubcaption{0.0, 0.0}{\label{fig:wp1k}}
        \oversubcaption{0.5, 0.0}{\label{fig:wp1x}}
        \oversubcaption{0.0, 0.5}{\label{fig:wp2k}}
        \oversubcaption{0.5, 0.5}{\label{fig:wp2x}}
    \end{subcaptions}
    \caption{\label{fig:wavepacket}%
        Two $m=0.8$ Gaussian meson wave packets, one per figure row. Specifically, absolute square of the coefficients of the fermionic field operators in Eq. (20), in their (a),(c) momentum and (b),(d) position space representation.
        The wave packet parameters are reported in the following table:
    }

    \begin{center}
        \begin{tabular}{cCCCCCC}\toprule
                                               & \kmean & \ksdev & \dkmean & \dksdev & \xmean & \dxmean \\\midrule
            \sref{fig:wp1k}, \sref{fig:wp1x}\; & +0.81  & 0.11   & 0       & 0.09    & 60.5   & +1      \\
            \sref{fig:wp2k}, \sref{fig:wp2x}\; & -0.81  & 0.11   & 0       & 0.09    & 139.5  & -1      \\\bottomrule
        \end{tabular}
    \end{center}

    \justify
    The mean values in the table are highlighted by straight lines in the plots (blue for $\kmean=\ev{p+q}$ and $\xmean=\ev{y+z}$, magenta for $\dkmean=\ev{q-p}$ and $\dxmean=\ev{z-y}$).
    The dashed ellipses represent the threshold ($\abs*{\wpxx[2;3]} = 10^{-6}$) below which we truncate the position space amplitude when constructing the MPO in \cref{fig:meson_MPO}.
\end{figure}
Meson wave packets are prepared starting from the \qed{} ground state $\ket{\qedvac}$ and acting with operators
\begin{equation}
    \begin{aligned}\label{eq:meson_wavepacket_creation}
        \meswp*
         & = \sumk \wpk \mesk*
        = \kspc^2\stgsumk*[2;3] \wpkk[2;3] \posk{\dress}[2]* \negk{\dress}[3]*                                                   \\
         & = \spc*^2\sumx*[2;3] \wpxx[2;3] \stgx[2]* \stgx[3] \prod_{\mathclap{\xx\ge\xx[2],\,\xx'\ge\xx[3]}} \cmpR* \cmpR[\xx']
        \,,
    \end{aligned}
\end{equation}
with \begin{math}
    \wpkk[2;3] = \wpk[\kk[2]+\kk[3]] \meswf[\kk[3] - \kk[2]]
\end{math}.
As for the free theory, $\wpxx[2;3]$ is implicitly defined in terms of $\wpkk[2;3]$ via \cref{eq:ab_def}.
The functions $\wpkk[2;3]$ and $\wpxx[2;3]$ for a pair of Gaussian meson wave packets, with $\wpk$ given by \cref{eq:gaussian_wavepacket}, are plotted in \cref{fig:wavepacket}.
The states prepared with $\meswp*$ operators are normalized a posteriori because, in our approximation, no exact orthonormality condition analogous to \cref{eq:fock_orthonormal} holds a priori for the states created by $\mesk*$, when $\cpl>0$.
\begin{figure}
    \begin{subcaptions}{\includegraphics{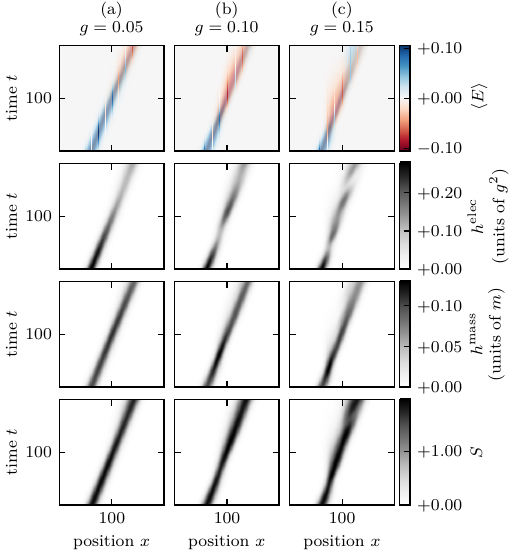}}
        \oversubcaption{0.0, 0.0}{\label{fig:single_meson_0.05}}
        \oversubcaption{0.0, 0.0}{\label{fig:single_meson_0.10}}
        \oversubcaption{0.0, 0.0}{\label{fig:single_meson_0.15}}
    \end{subcaptions}
    \caption{\label{fig:single_meson}%
        Test of the stability of \qed{} mesons prepared using \cref{eq:mesonwf,eq:meson_wavepacket_creation}.
        The electric field $\ev*{\eleR}$, its energy density $\helec_{\xx}$, the mass energy density $\hmass_{\xx}$ and the entanglement entropy $\ent_{\xx}$ (rows) are measured during the propagation of a single meson, for mass $\mass=0.8$ and increasing values of the coupling $\cpl$ (columns).
    }
\end{figure}
\begin{figure}
    \includegraphics{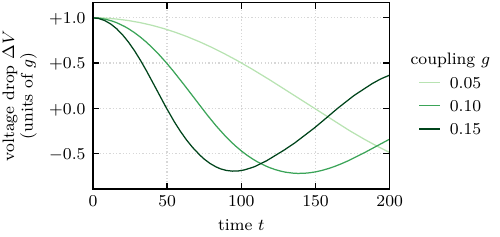}
    \caption{\label{fig:polarization(t)}%
        Damped oscillation of the voltage drop across the chain $\Vdrop$ during the propagation of the a single meson (simulations in \cref{fig:single_meson}) for different values of the coupling $\cpl$.
        Notice the increase of the oscillation frequency with the coupling $\cpl$.
    }
\end{figure}
\begin{figure*}[t]
    \begin{subcaptions}{\includegraphics{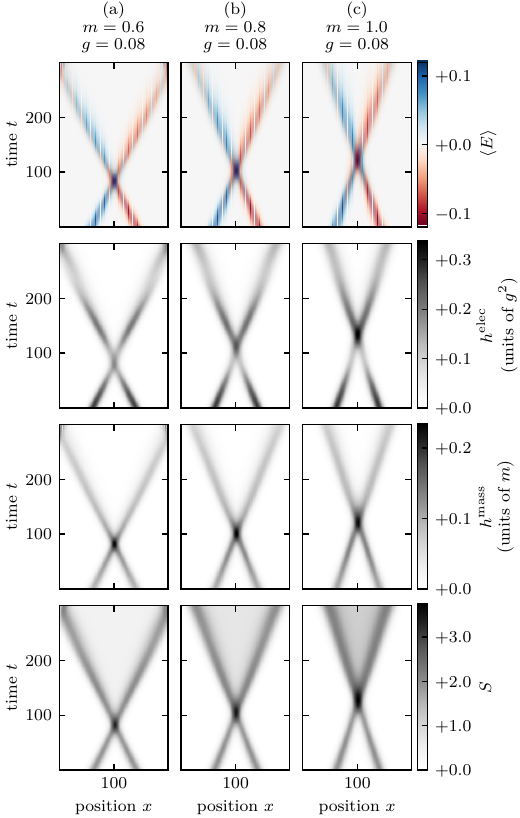}}
        \oversubcaption{0.0, 0.0}{\label{fig:dynamics_m}}
        \oversubcaption{0.0, 0.0}{\label{fig:dynamics_mm}}
        \oversubcaption{0.0, 0.0}{\label{fig:dynamics_mmm}}
    \end{subcaptions}
    \hfill
    \begin{subcaptions}{\includegraphics{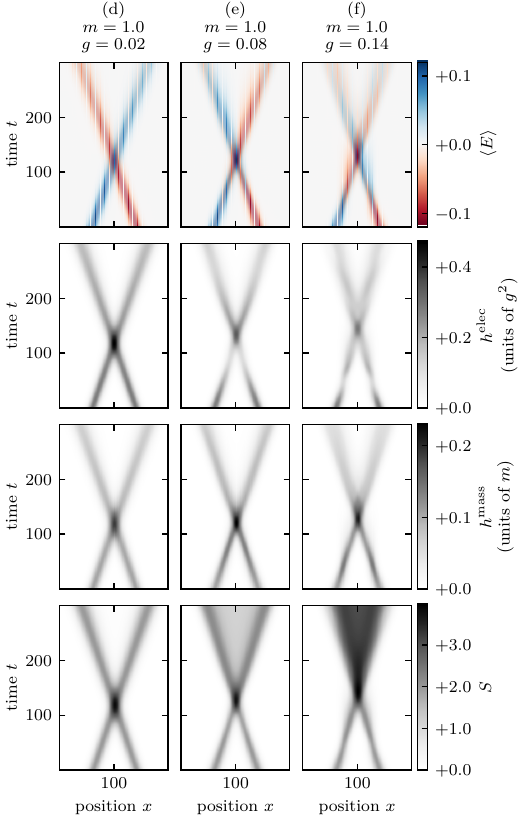}}
        \oversubcaption{0.0, 0.0}{\label{fig:dynamics_g}}
        \oversubcaption{0.0, 0.0}{\label{fig:dynamics_gg}}
        \oversubcaption{0.0, 0.0}{\label{fig:dynamics_ggg}}
    \end{subcaptions}
    \caption{\label{fig:dynamics}%
        Simulations of meson-meson collisions in \qed{}.
        Measurements of the electric field $\ev*{\eleR}_t$, its energy density $\helec_{\xx}(t)$, the mass energy density $\hmass_{\xx}(t)$ and the entanglement entropy $\ent(t, x)$ are reported (rows).
        Various values of the mass $\mass$ (left block) and coupling $\cpl$ (right block) parameters are considered.
        The initial meson wave packet distributions are those of \cref{fig:wavepacket}.
    }
\end{figure*}
Before focusing on meson-meson scatterings, we simulate the free propagation of one meson and test the approximation in \cref{eq:mesonwf} in the parameters region explored hereafter.
The time evolution the meson wave packet in \cref{fig:wp1k,fig:wp1x} is shown in \cref{fig:single_meson} for mass $\mass=0.8$ and couplings up to $\cpl = 0.15$.
Similar results are observed for masses $\mass=0.6$ and $\mass=1.0$.
As anticipated, the meson does not decay during the simulations and, crucially, its entanglement track remains confined in the region where the wave packet is localized.
On the other hand, we do observe some internal dynamics in the wave packets, taking the form of periodic inversions of the meson polarization (sign of its electric string), whose frequency increases with the coupling $\cpl$ \cite{Pichler2016RealTimeDynamics,Surace2020LatticeGaugeTheories}.
This behavior is also captured by \cref{fig:polarization(t)}, where the voltage drop across the chain,
\begin{math}
    \Vdrop = \sumx \cpl\eleR
\end{math},
is plotted as a function of time.
Looking at the relevant rows of \cref{fig:single_meson} we see that the inversions are accompanied by a drop in the electric field energy $\helec$ and a concentration of the mass energy density $\hmass$.
We infer that the inversions correspond to damped oscillations of the fermion and antifermion constituents of the meson around their center of mass.

As far as the interpretation of the results is concerned, hereafter we consider as exact the ansatz for the meson creation operator in \cref{eq:mesonwf}.
For instance, in the analysis of \cref{sec:entanglement} we assume $\mesk*$ creates monochromatic mesons,  neglecting the consequences of the internal dynamics on the entanglement.

\section{Dynamics}\label{sec:dynamics}
In this Section we present some tensor network simulations of lattice \qed{} meson-meson scatterings.
We explore a region of model parameters ranging from weak ($\cpl \ll \mass$) to intermediate ($\cpl / \mass \approx 1/4$) coupling.
All the reported simulations follow the scheme outlined in \cref{sec:methods}.
They are carried out in the center of mass frame of reference and for parity-symmetric initial configurations.
Accordingly, the initial scattering state is prepared by using a pair of uncorrelated meson wave packet creation operators form \cref{eq:mesonwf,eq:meson_wavepacket_creation} with identical $\ksdev$ and $\dksdev$, opposite mean momenta $\kmean$, and symmetric $\xmean$ and $\dxmean$.

\subsection{Scattering phenomenology}
\begin{figure}
    \includegraphics{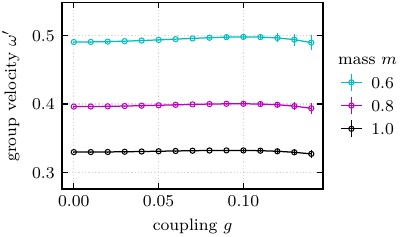}
    \caption{\label{fig:group_velocity(g)}%
        Mass $\mass$ and coupling $\cpl$ dependence of the meson group velocity $\pulse[]'$, as defined in \cref{eq:group_velocity_fit} and measured during the first (approximately free) stage of the propagation.
        Error bars represent three standard deviations.
        The uncertainty increases with the coupling $\cpl$ because stronger interactions anticipate the deflection of the meson trajectories.
    }
\end{figure}

An overview of the dynamics is shown in \cref{fig:dynamics}.
\Cref{fig:dynamics_m,fig:dynamics_mm,fig:dynamics_mmm} show the scattering of two initial mesons
for increasing bare mass $\mass=0.6,0.8,1.0$ and fixed coupling strength $\cpl=0.08$.
Increasing $\mass$ clearly affects the kinematics by slowing down the propagation of the mesons before and after the collision but it also results in the generation of a larger entanglement between the scattering products.
To quantify the first effect we focus on the initial stage of the evolution ($t<30$) and linearly interpolate the trajectories $\xmean(t)$ of the mass energy density peaks:
\begin{equation}\label{eq:group_velocity_fit}
    \xmean(t) = \pulse[]' t + \xmean(0)
    \,.
\end{equation}
The resulting group velocities $\pulse[]'$ are plotted in \cref{fig:group_velocity(g)}.
The second effect, namely the increased entanglement after the scattering, may be interpreted as an indirect consequence of the slow down of the  colliding particles, as they effectively interact for a longer time.

\Cref{fig:dynamics_g,fig:dynamics_gg,fig:dynamics_ggg} show again the scattering of the initial mesons for fixed bare mass $\mass=1.0$ and coupling $\cpl=0.02,0.08,0.14$.
We observe a drastic increase of the post-collision entanglement with the strength $\cpl$ of the interactions.
A detailed discussion of this phenomenon is given in \cref{sec:entanglement}.
Furthermore, in \cref{fig:dynamics_g,fig:dynamics_gg} the polarizations of the outgoing mesons are inverted, as indicated by the sign of the electric field.
The string inversion, however, is not a consequence of the collision.
As discussed in \cref{sec:initial_state}, inversions are an (accidental) internal feature of our meson states and are observed also in \cref{fig:single_meson}, during the free propagation of a single meson.

\subsubsection{Energy balance}
\begin{figure}
    \begin{subcaptions}{\includegraphics{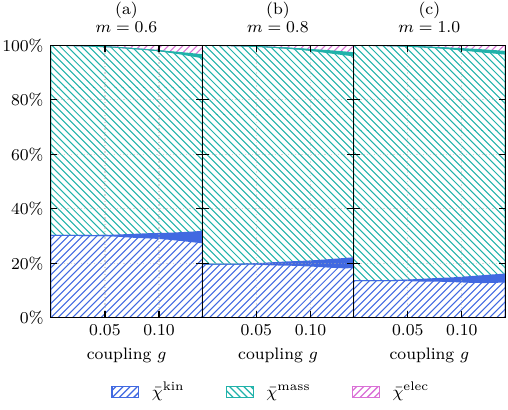}}
        \oversubcaption{0.05, 0.05}{\label{fig:energy_balance_l}}
        \oversubcaption{0.30, 0.05}{\label{fig:energy_balance_m}}
        \oversubcaption{0.55, 0.05}{\label{fig:energy_balance_h}}
    \end{subcaptions}
    \caption{\label{fig:energy_balance}%
        Total energy partition in the mass, kinetic and electric energy contributions from \cref{eq:energy_fractions}: time averages (hatched regions) and fluctuations (filled regions) during various meson-meson scatterings.
        Masses $\mass=0.6$ \sref{fig:energy_balance_l}, $0.8$ \sref{fig:energy_balance_m} and $1.0$ \sref{fig:energy_balance_h} and couplings ranging from $\cpl=0$ to $\cpl=0.14$ are considered.
        The thickness of the filled regions reproduce a $\pm2\epsilon^\gamma$ confidence belt around the mean value $\hfrac^\gamma$ of the kinetic and mass energy fractions.
    }
\end{figure}
The total energy $\ev*{H}$ of a \qed{} state can be partitioned in mass, electric and kinetic energy fractions $\hfrac^{\gamma}$, $\gamma\in\{\text{mass},\text{kin},\text{elec}\}$.
In the simulation frame of reference these are defined as
\begin{equation}\label{eq:energy_fractions}
    \begin{gathered}
        \fracmass = \frac{\sumx\hmass_{\xx}}{\ev*{H}}
        \,, \quad
        \fracelec = \frac{\sumx\helec_{\xx}}{\ev*{H}}
        \,, \\[1ex]
        \frackin = 1 - \fracmass - \fracelec
        \,.
    \end{gathered}
\end{equation}
We monitor $\hfrac^{\gamma}(t)$ during various meson-meson scatterings.
The time averages over the whole simulation time span, denoted $\bar{\hfrac}^{\gamma}$, are reported in the stacked area plots of \cref{fig:energy_balance}, together with the standard deviations (time fluctuations) $\epsilon^{\gamma}$,
\begin{math}
    (\epsilon^{\gamma})^2 = \overline{(\hfrac^{\gamma}(t) - \bar{\hfrac}^{\gamma})^2}
\end{math},
of the mass and kinetic fractions.
The mass and kinetic energies are the dominant contributions in the parameter region we explored, while the electric energy fraction never exceeds $5\%$, decreasing mildly with the mass and growing as $\cpl^2$ for large enough couplings ($\cpl \ge  0.05$).
The relative weight of the mass and kinetic energies is mostly controlled by the mass parameter.
On the other hand, energy transfers (namely, amount of fluctuations $\epsilon^{\gamma}$) are boosted for larger couplings, i.e., stronger interactions.
We stress that, even for the largest simulated couplings, the mass energy is approximately constant during the evolution, with time fluctuations amounting at most to $0.5\%$ of its value.
This behavior strongly hints that the simulated processes are elastic collisions.
In $(1+1)$-dimensions, energy-momentum conservation implies that the products of an elastic collision of two particles (of equal mass) have the same momenta of the incoming particles.
In the next Section we verify this kinematical constraint by analyzing the momentum content of the initial and final states.

\subsubsection{Momentum space analysis}

Among the most important observables for a scattering process are the species and the momenta of the  incoming and outgoing  particles involved in the collision.
In experiments, the momenta of the incoming particles are tuned by collimating the colliding particle beams.
The species and momenta of the scattering products, instead, are inferred from detector data.
Our simulations follow a similar procedure:
the momenta of the incoming particles are set by the initial state, while those of the scattering products are identified analyzing the final state.
The momenta of the excitations contained in a state are reflected by its spatial periodicities.
To detect the correlations between pairs of meson excitations in different positions we evaluate the connected $2$-point correlation function for the meson composite operator
\begin{math}
    \mesx = \stgx\stgx[\xx+\spc]*
\end{math}
($\xx\in\sys{E}$), namely
\begin{math}
    \left.
    \corrx = \ev*{\mesx[2]\mesx[3]*} - \ev*{\mesx[2]}\hspace{-1ex}\ev*{\mesx[3]*}
    \right.
\end{math}.
Then, we compute the translation invariant momentum space connected $2$-point function
\begin{equation}\label{eq:correlator}
    \corrk = \frac{(2\spc*)^2}{2\pi}\!\sumx*[\xx[2;3]\in\sys{E}] e^{-i\kk(\xx[2]-\xx[3])} \corrx
    \,.
\end{equation}

\begin{figure}
    \begin{subcaptions}{\includegraphics{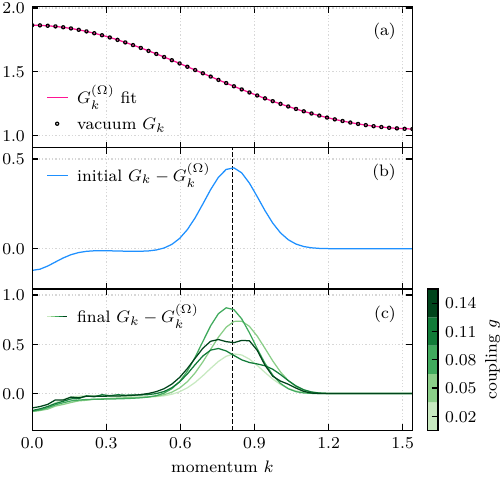}}
        \oversubcaption{0.0, 0.0}{\label{fig:correlators_vac}}
        \oversubcaption{0.0, 0.3}{\label{fig:correlators_in}}
        \oversubcaption{0.0, 0.6}{\label{fig:correlators_out}}
    \end{subcaptions}
    \caption{\label{fig:correlators}%
        Time-independent momentum space connected meson-meson correlation functions, $\corrk$, at $\mass=0.6$ and for positive momenta $\kk$ (the $\kk<0$ branch is symmetric).
        Evaluated:
        at $\cpl=0.08$ on the full \qed{} vacuum \sref{fig:correlators_vac}
        and the initial meson-meson scattering state \sref{fig:correlators_in};
        as well as on final scattering states for various couplings $\cpl$ (see color bar) \sref{fig:correlators_out}.
        The vacuum and initial $\corrk$ are almost coupling $\cpl$ independent (up to percent order deviations).
        In \sref{fig:correlators_vac} the fitted $\corrkvac$ from \cref{eq:correlator_vacuum} is also reported, while in \sref{fig:correlators_in} and \sref{fig:correlators_out} it has been subtracted.
        The dashed vertical line corresponds to the mean momentum of the (left) initial wave packet.
    }
\end{figure}
Before evaluating \cref{eq:correlator} on the initial and final scattering states of our simulations, let us discuss the case of the ground state.
Due to the nonzero correlation length $\corrlen$ of the full \qed{} vacuum, in the thermodynamic limit \cite{CorrelatorDecayNote},
\begin{equation}\label{eq:correlator_vacuum}
    \corrxvac \propto e^{-\abs{\xx[2] - \xx[3]}/\corrlen}
    \,,\quad
    \corrkvac \propto \frac{\sinh(2\spc*/\corrlen)}{\cosh(2\spc*/\corrlen) - \cos(2\spc*\kk)}
    \,.
\end{equation}
Fitting the numerical $\corrk$ via \cref{eq:correlator_vacuum} we find that the thermodynamic limit provides an excellent approximation of the numerical results (up to a constant shift) see \cref{fig:correlators_vac}.
Moreover, for all the simulated $\mass$ and $\cpl$ values we find
\begin{math}
    \corrlen / (\spc\cells) \approx \order*{10^{-3}} \ll 1
\end{math};
signaling that we are indeed at the thermodynamic limit and justifying a posteriori the usage of the expressions in \cref{eq:correlator_vacuum,eq:correlator}.

The 2-point function $\corrk$ of initial and final scattering states also presents a background of the type in \cref{eq:correlator_vacuum} (up to a shift).
On top of this background, we observe peaks detecting the momenta of the incoming and outgoing mesons.
The momentum space correlators of some initial and final scattering states are plotted in \cref{fig:correlators_in,fig:correlators_out}, with the background removed.
All initial and final correlators are peaked around the mean momentum of the initial wave packet, as expected for a $2 \to 2$ elastic scattering in $1+1$ dimensions.
The distortions appearing in final state correlators are likely caused by the inexact modelling of the meson excitations.

\subsection{Towards \boldmath\texorpdfstring{$S$}{S}-matrix elements}

The central quantity in scattering theory is the $S$ operator or \smatrix{} \cite{Weinberg1995QuantumTheoryFields}.
In this Section, after having briefly introduced the problem in the continuum, we lay out a prescription for extracting \smatrix{} elements from dynamical lattice simulations.

\smatrix{} elements are essential to verify the predictions of a theoretical model, as they relate the particle content of the initial (infinite past) and final (infinite future) states of a scattering experiment \cite{Weinberg1995QuantumTheoryFields}.
Here we set coordinates in which the collision takes place at $x=t=0$.
Ideally, the \smatrix{} reads
\begin{math}
    S = \lim_{t\to\infty}e^{-2itH}
\end{math},
where $H$ is the Hamiltonian of the model.
However, this limit involves infinitely oscillating phases and thus does not exist; nor does
\begin{math}
    \lim_{t\to\pm\infty}\ket{\P(t)}
\end{math},
where
\begin{math}
    \ket{\P(t)} = e^{-itH}\ket{\P}
\end{math}
describes the state of a system undergoing a scattering process.
To overcome this problem, the asymptotic evolution has to be factored out in the definition of the \smatrix{} \cite{Hannesdottir2020SMatrixMassless,Strocchi2013IntroductionNonPerturbative}.
Explicitly,
\begin{equation}\label{eq:S-matrix}
    S = \Omega_+^\dagger \Omega_-^\phdag
    \,, \quad
    \Omega_\pm = \lim_{\ \mathclap{t\to\pm\infty}\ } e^{itH}e^{-itH_0}
    \,;
\end{equation}
where $H_0$ is the Hamiltonian describing the free kinematics of the stable particle states of the theory defined by $H$.
The definition in \cref{eq:S-matrix} is motivated by the assumption \cite{Buchholz2005ScatteringRelativisticQuantum,Hannesdottir2020SMatrixMassless,Strocchi2013IntroductionNonPerturbative,AsymptoticNote} that the interaction decays rapidly enough so that, at asymptotic times $t\to\pm\infty$, when particles are far apart,
the evolution specified by $H$ coincides with that of $H_0$ and the scattering solution $\ket{\P(t)}$ approaches the trajectories of some freely evolving particle states
$\ket{\Pf_\pm}$; namely
\begin{math}
    e^{-itH}\ket{\P} \sim e^{-itH_0}\ket{\Pf_{\pm}}
\end{math}.
Parametrizing a complete set of asymptotic configurations as $\{\ket{\Pf_\alpha}\}$, the \smatrix{} elements read
\begin{math}
    S_{\alpha'\hspace{-1pt}\alpha} = \mel{\Pf_{\alpha'}}{S}{\Pf_{\alpha}}
\end{math}.
The index $\alpha$ typically runs over momenta, spin projections and possibly other discrete labels \cite{Weinberg1995QuantumTheoryFields}.

Two observations are in order.
As just mentioned, \smatrix{} elements are usually specified for momenta (and energy) eigenstates.
However, these states are completely delocalized in space (and time) and thus cannot describe noninteracting particles localized in far-distant space regions.
Indeed, a limit of narrow momentum space wave packets is implied in the previous construction \cite{Strocchi2013IntroductionNonPerturbative,Weinberg1995QuantumTheoryFields}.
Moreover, the infinite time limits are a useful idealization \cite{Buchholz2005ScatteringRelativisticQuantum,Weinberg1995QuantumTheoryFields}:
in real world experiments, measurements are carried out at macroscopic times that precede and follow the collision by a time lapse $\tf$,
much larger than the microscopic timescale of the collision itself but still finite.
These measurements should reveal a state that approximates a free multiparticle wave packet with a degree of precision related to that of the time limit.
It follows that the transition amplitudes measured experimentally are
\begin{equation}\label{eq:transitions}
    \ampl(\Pf\to\Pf') = \mel{\Pf'(+\tf)}{e^{-2i\tf H}}{\Pf(-\tf)}
    \,,
\end{equation}
with $\ket{\Pf(-\tf)}$ and $\ket{\Pf'(+\tf)}$ wave packets of approximate free multiparticle energy-momentum eigenstates.

With tensor network scattering simulations, the evaluation of the transition amplitudes in \cref{eq:transitions} is straightforward.
To this aim, the initial state $\ket{\Pf(-\tf)}$ is evolved until some scattering products emerge, after the collision, in the form of well separated particle wave packets.
An estimate of how well the evolving state resembles a state of isolated particles is obtained, e.g., inspecting its mass energy density profile, in particular the location and width of its peaks.
Once the desired separation is reached, the state $\ket{\Pf(\tf)}=e^{-2i\tf H}\ket{\Pf(-\tf)}$ is stored and the simulation is terminated.
At this point, \cref{eq:transitions} provides the amplitude of the transition\emdash{}due to the collision\emdash{}from the initial state, $\ket{\Pf(-\tf)}$, to any $\ket{\Pf'(+\tf)}$ we are interested in.

\begin{figure}
    \begin{subcaptions}{\includegraphics{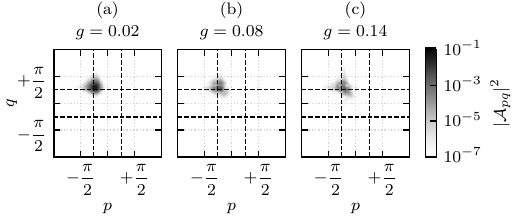}}
        \oversubcaption{0.05, 0.05}{\label{fig:transitions_w}}
        \oversubcaption{0.30, 0.05}{\label{fig:transitions_m}}
        \oversubcaption{0.55, 0.05}{\label{fig:transitions_s}}
    \end{subcaptions}
    \caption{\label{fig:transitions}%
        Transition probabilities $\sabs{\ampl_{\kL\kR}}$ to the states defined in \cref{eq:LR_amplitude}, for three meson-meson scattering simulations with bare mass $\mass=0.8$ and coupling $\cpl=0.02$ \sref{fig:transitions_w}, $0.08$ \sref{fig:transitions_m}, and $0.14$ \sref{fig:transitions_s}.
        Namely, probability of the two mesons in \cref{fig:wavepacket} to evolve, after a collision, into pairs of meson wave packets peaked around momenta $\kk[2]$ and $\kk[3]$ and delocalized in the left and right side of the chain, respectively.
        The dashed lines correspond to the initial meson momenta $\kmean\approx\pm\pi/4$.
        The resolution of the above images is related to the momentum space standard deviation $\sharpksdev\approx0.14$ of these wave packets and thus, indirectly, to the lattice size.
        The transition probabilities in Figure are computed projecting on a family of final states with $\kk[2]$ and $\kk[3]$ values spaced by $\sharpksdev/2$.
    }
\end{figure}
Here we identify $\tf$ with the time at which the mass energy density peaks of the outgoing wave packets are separated by $\Delta x \gtrsim 100$ and focus on transitions amplitudes to final states $\ket{\Pf'(+\tf)}$ of two mesons.
In order to study the distribution of their momenta, we consider meson wave packets with amplitudes $\wpLk_{\kk}$ and $\wpRk_{\kk}$ peaked at momenta $\kL$ and $\kR$ and completely delocalized in the left and right half of the chain respectively.
That is, we compute the overlap of the final state with states of the form
\begin{equation}\label{eq:LR_amplitude}
    \ket{\Pf'_{\kL\kR}(\tf)} =
    \normalization_{\kL\kR}
    \kspc^2\sumk*[\kk'\kk] \wpRk_{\kk'}\wpLk_{\kk}
    \mesk[{\kk'}]*\mesk*
    \ket{\qedvac}
    \,,
\end{equation}
with $\normalization_{\kk[2;3]}$ enforcing the normalization and
\begin{equation}\label{eq:R_wavepacket}
    \wpRk_{\kk} = \fouriercoef\,\sumx e^{-i(\kk-\kR)\xx}\,\Theta(\xx)
    \,,
\end{equation}
$\Theta(\xx)$ being the step function.
Similarly for $\wpLk$.
Exemplary transition probabilities
\begin{math}
    \sabs{\ampl_{\kL\kR}}=\sabs*{\braket{\Pf'_{\kL\kR}(\tf)}{\Pf(\tf)}}
\end{math}
are plotted in \cref{fig:transitions}.
As for \cref{fig:correlators} and as expected for kinematical reasons, the momentum distributions of the final meson are concentrated exactly on the momentum space support of the initial wave packets.

As a final remark, let us stress that it is in principle possible to extract the \smatrix{} defined in \cref{eq:S-matrix} from lattice tensor network simulations.
The starting point are the finite time amplitudes in \cref{eq:transitions}:
in order to correctly identify which \smatrix{} element is computed, the free $H_0$ evolution of the wave packets $\ket{\Pf'(+\tf)}$ and $\ket{\Pf(-\tf)}$ from time $t = 0$ to $t = \pm \tf$ has to be compensated in \cref{eq:transitions}.
In this way approximate (due to the finite $\tf$) and smeared (due to the finite momentum spread of the wave packets) \smatrix{} elements are obtained.
Finally,
the continuum and thermodynamic limits have to be performed.
The thermodynamic limit reads $\cells,\tf\to\infty$, that is, the limit is approached not only increasing the lattice size ($\spc\cells$), but also, simultaneously, the duration of the simulation ($\tf$), while keeping the reference wave packet, $\ket{\Pf'(0)}$ and $\ket{\Pf(0)}$, unchanged.
In the $\tf\to\infty$ limit, $\ket{\Pf'(+\tf)}$ and $\ket{\Pf(-\tf)}$ approach states of truly noninteracting particles.
\smatrix{} elements between initial and final states of definite momentum are then extrapolated taking a limit of vanishing momentum spread of the reference wave packets.
In \cref{app:heaviside_wavepacket} we show that
the wave packets $\wpLk$ and $\wpRk$ behave as momentum projectors in the continuum and thermodynamic limit.

\section{Entanglement Generation}\label{sec:entanglement}
The Von Neumann entanglement entropy $\ent$ observed in the scattering simulations of \cref{sec:dynamics} can be traced back to three major sources: the ground state ($\Sgrn$, background), the particle\emdash{}either fermion, antifermion or meson\emdash{}wave packets ($\Swps$, intraparticle entanglement) and their interactions ($\Sint$, interparticle entanglement).
As a first approximation, we consider these contributions as additive.

The models in \cref{sec:models} are translationally invariant in the thermodynamic limit or for periodic boundaries.
Hence, the ground state $\ket{\vac}$ is responsible for a uniform (up to boundary effects) background entanglement $\Sgrn$.
In what follows we estimate a priori
the wave packet contribution to infer the interaction entanglement generated by the dynamics.

\subsection{Wave packet entanglement}

Acting on $\ket{\vac}$ with the wave packet creation operators from \cref{eq:free_wavepacket_creation,eq:meson_wavepacket_creation}, additional entanglement, $\Swps$, appears on top of $\Sgrn$ in the space region where the particles are localized.
This entanglement contribution can be characterized for freely propagating wave packets in the $\mass\to\infty$ limit, assuming that the contributions of different wave packets are additive, as expected for uncorrelated wave packets.
We checked this assumption for a few overlapping wave packets of either fermions or antifermions, and found additivity violations (due to the exclusion principle) of a few percent order.

\subsubsection{Fermion and antifermion entanglement}
We consider a single fermion wave packet
\begin{math}
    \ket{\P}=\smash{\poswp*}\ket{\barevac}
\end{math},
prepared via \cref{eq:fermion_wavepacket_creation};
analogous results hold for an antifermion.
In the infinite mass limit, the Hamiltonian in \cref{eq:Hfree_x} is dominated by the mass term
and the ground state approaches the bare vacuum\emdash{}i.e., the Néel product state
\begin{math}
    \ket{\barevac} = \ket{0101\ldots}
\end{math}.
Consider a bipartition of the chain $\latx$ in the subsystems
\begin{math}
    \sys{L} = \{ \xx[2]<\xx \}
\end{math}
and
\begin{math}
    \sys{R} = \{ \xx[2]\ge\xx \}
\end{math}
with $\xx\in\sys{E}$ (even sublattice), so that
\begin{math}
    \ket{\barevac} = \subket{L}{\barevac}*\otimes\subket{R}{\barevac}
\end{math}
and $\ket{\P}$ decomposes as
\begin{align}
    \ket{\P}
     & =
    \Big[ \sumx[\xx[2]\in\sys{L}] \wpx[2]^{\poswpSym} \stgx[2]* \subket{L}{\barevac}* \Big]
    \otimes \subket{R}{\barevac}*
    +
    \subket{L}{\barevac}* \otimes
    \Big[ \sumx[\xx[2]\in\sys{R}] \wpx[2]^{\poswpSym} \stgx[2]* \subket{R}{\barevac}* \Big]
    \nonumber
    \\
     & =
    \sqrt{\smash[b]{\cdf}} \subket{L}{\P}* \otimes \subket{R}{\barevac}* +
    \sqrt{\smash[b]{1-\cdf}} \subket{L}{\barevac}* \otimes \subket{R}{\P}*
    \label{eq:fermion_wavepacket_decomposition}
    \,,
\end{align}
with
\begin{equation}
    \subbraket{S}{\P}{\P}* = 1
    \,,\qquad
    \subbraket{S}{\P}{\barevac}* = 0
    \,,
\end{equation}
where
\begin{math}
    \sys{S}\in\{\sys{L},\sys{R}\}
\end{math}.
Furthermore, exploiting the expression of $\wpx^{\poswpSym}$ in \cref{eq:wpx_limit}, $\cdf$ is identified with the cumulative distribution function (CDF)
\begin{math}
    \cdf = \sumx[\xx[2]<\xx] \sabs*{\wpx[2]^{\poswpSym}}
\end{math}.
\Cref{eq:fermion_wavepacket_decomposition} represents the coherent superposition of the overall fermion excitations in each side of the chain.
In a basis obtained extending
\begin{math}
    \{\subket{S}{\barevac}*,\subket{S}{\P}\}
\end{math}
to a complete orthonormal system, the reduced density matrix of the $\sys{S}$ subsystem reads
\begin{math}
    \rho = \diag(\cdf, 1-\cdf, 0, 0, \ldots)
\end{math},
yielding the entanglement entropy profile
\begin{equation}\label{eq:wavepacket_entanglement}
    \ent(\xx) = - \cdf \log_2 \cdf - (1-\cdf) \log_2 (1-\cdf)
    \,.
\end{equation}
Note how, at the median $\tilde{x}$ of the wave packet distribution $\cdf[\tilde{\xx}]=1/2$ and $\ket{\P}$ in \cref{eq:fermion_wavepacket_decomposition} becomes a Bell state, thus $\ent(\tilde{\xx})=1$.

Specializing to a Gaussian (fermion or antifermion) wave packet,
it follows from \cref{eq:free_cdf_limlim} that in the thermodynamic and continuum limits $\cdf$ becomes
the CDF of a Gaussian distribution with mean $\xmean$ and standard deviation $1/2\ksdev$:
\begin{equation}\label{eq:normal_cdf}
    \cdf = \frac{1}{2}\left[ 1 + \erf\left(\frac{x-\xmean}{\sqrt{2}\xsdev}\right) \right]
    \,,\quad
    \xsdev=1/2\ksdev
    \,.
\end{equation}
An analytic expression for $\ent(\xx)$, denoted $\Sgauss(\xx; \xmean, \xsdev)$, is obtained substituting $\cdf$ from \cref{eq:normal_cdf} in \cref{eq:wavepacket_entanglement}.
As shown by \cref{fig:Swps(xt)},
$\Sgauss(\xx; \xmean, \xsdev)$ closely resembles the associated probability density function (PDF)
\begin{math}
    \pdf \propto e^{-(\xx-\xmean)^2/2\xsdev^2}
\end{math}
(up to the normalization), even though it  decays as
\begin{math}
    \xx e^{-(\xx-\xmean)^2/2\xsdev^2}
\end{math} and thus has heavier tails.

\subsubsection{Meson entanglement}
The previous results are straightforwardly extended to meson wave packets exploiting the formulation of \qed{} with the link degrees of freedom integrated out from \cref{eq:Hqed_integrated}.
In the infinite mass limit the ground state is again the Néel state.
Suppose that the meson wave packet creation operator in \cref{eq:meson_wavepacket_creation} factorizes as a product of fermion and antifermion creation operators,
\begin{equation}\label{eq:meson_wavepacekt_creation_factorization}
    \meswp* = \poswp[\wpSym^+]*\negwp[\wpSym^-]*
    \,,
\end{equation}
for some amplitudes $\wpSym^{\pm}$.
Then, by the additivity assumption, the entanglement profile of the meson is the sum of a pair of \cref{eq:wavepacket_entanglement} contributions,
\begin{math}
    \ent = \ent_{\wpSym^+} + \ent_{\wpSym^-}
\end{math}.
\Cref{fig:wavepacket} shows that this is a good approximation for the Gaussian mesons of our simulations, which are characterized by $\ksdev\approx\dksdev$.
Indeed, in this case, \cref{eq:meson_wavepacekt_creation_factorization} holds up to $\order*{\ksdev^2-\dksdev^2}$ terms, with $\wpSym^{\pm}_{\kk}$ Gaussian amplitudes whose parameters are related to those of the meson wave packet by the substitutions
\begin{equation}
    \kmean \to \frac{\kmean}{2}
    ,\;\;
    \ksdev \to \frac{\sqrt{\ksdev^2+\dksdev^2}}{2}
    ,\;\;
    \xmean \to \xmean\mp\frac{\dxmean}{2}
    .
\end{equation}
Moreover, the ${\dxmean}/{2}$ phase shift can be neglected if the fermion and antifermion are almost overlapped, i.e., if $\dxmean \ll 1/\ksdev$, as is the case for the mesons in \cref{fig:wavepacket}.

We conclude that the entanglement due to a Gaussian meson wave packets is approximated by
\begin{math}
    2\Sgauss(\xx;\xmean,\xsdev)
\end{math},
where $\Sgauss$ is the result derived for fermions and antifermions and
\begin{math}
    \xsdev
    = 1/\sqrt{\ksdev^2+\dksdev^2}
\end{math}.

\subsubsection{Entanglement propagation}

The evolution of a particle wave packet
\begin{math}
    \sumk*\wpk\ket{\kk}
\end{math},
i.e., a wave packet of eigenstates $\ket{\kk}$ of momentum $\kk$ and energy $\pulse$, is given by
\begin{math}
    \wpk(t) = e^{-i\pulse t} \wpk
\end{math}.
This is true for fermions and antifermions in the free theory, as well as for (exact) mesons in \qed{}, even though the dispersion relation $\pulse$ of mesons is not known analytically if $\cpl>0$.

Let us focus again on a Gaussian wave packet in the continuum with $\wpk(t=t_0)$ given by \cref{eq:gaussian_wavepacket}.
A stationary-phase approximation shows that the absolute square of the inverse Fourier transform of $\wpk(t)$ is a Gaussian PDF also for $t \neq t_0$, but translated and widened according to
\begin{subequations}\label{eq:gaussian_evolution}
    \begin{align}
        \xmean(t)   & = \xmean - \pulse[\kmean]' (t-t_0)
        \,,                                                                                 \\
        \xsdev^2(t) & = \xsdev^2 + \bigg(\frac{\pulse[\kmean]''}{2\xsdev}\bigg)^2 (t-t_0)^2
        \,.
    \end{align}
\end{subequations}
Combined with \cref{eq:free_cdf_limlim} and the results of the previous paragraphs, \cref{eq:gaussian_evolution} provides the entanglement of freely propagating fermions, antifermions and (factorizable) mesons.
\begin{figure}
    \includegraphics{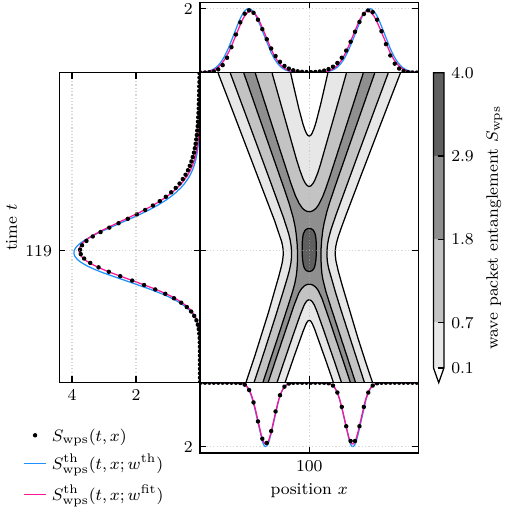}
    \caption{\label{fig:Swps(xt)}%
        Wave packet entanglement entropy during the free propagation of the mesons in \cref{fig:wavepacket} for $\mass=1$ and $\cpl=0$.
        Comparison of the numerical results with the $\Swpsth(t,x;\modpars)$ prediction from \cref{eq:wavepacket_entanglement_model}.
        The contour plot reproduces the numerical data $\Swps$.
        The bottom and top panels show the initial and final time spatial entanglement profiles; the left panel shows the midchain time profile.
        In each side panel, along with $\Swps$, we plot the relevant section of $\Swpsth(t,x;\modpars)$ for (blue lines) $\modpars=\modpars^{\text{th}}$ and (red lines) $\modpars=\modpars^{\text{fit}}$.
    }
\end{figure}
We model the wave packet entanglement due to a pair of parity-symmetric scattered mesons introducing
\begin{multline}\label{eq:wavepacket_entanglement_model}
    \Swpsth(t,x;\modpars) = \normalization \big[
        \Sgauss(x;\xmean(t),\xsdev(t))
        +\\
        \Sgauss(\cells-1-x;\xmean(t),\xsdev(t))
        \big]
    \,,
\end{multline}
where
\begin{math}
    \modpars = (\normalization\compeq2, t_0, \xmean, \xsdev, \pulse[\kmean]', \pulse[\kmean]'')
\end{math}
collects all the parameters entering \cref{eq:gaussian_evolution,eq:wavepacket_entanglement_model}.
Having been derived for freely propagating particles, \cref{eq:wavepacket_entanglement_model} applies only for $\cpl=0$ or when the mesons are distant enough so that their interaction can be neglected.
In \cref{fig:Swps(xt)} we compare the numerical wave packet entanglement $\Swps(t,x)$ from a $\cpl=0$ simulation with the prediction from $\Swpsth(t,x;\modpars)$, for both the expected parameters $\modpars=\thpars$, and the values $\fitpars$ that best interpolate the numerical data (setting $t_0=0$).
The function $\Swpsth$ models accurately the numerical entanglement;
comparing $\thpars$ and $\fitpars$, the prominent distortion is a slight squashing of the expected entanglement profiles due to the lack of exact additivity between the various entanglement contributions.

\subsection{Interaction entanglement}
\begin{figure}
    \begin{subcaptions}{\includegraphics{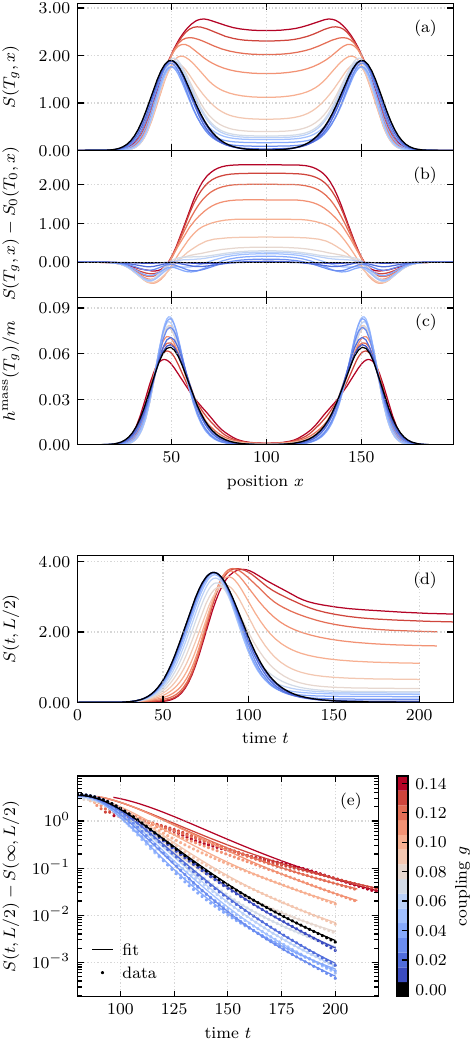}}
        \oversubcaption{0.1, 0.10}{\label{fig:S(x)}}
        \oversubcaption{0.1, 0.20}{\label{fig:Sint(x)}}
        \oversubcaption{0.1, 0.30}{\label{fig:pdf(x)}}
        \oversubcaption{0.1, 0.50}{\label{fig:S(t)}}
        \oversubcaption{0.1, 0.70}{\label{fig:Swps(t)}}
    \end{subcaptions}
    \caption{\label{fig:S(xt)}%
        Entanglement entropy $\ent$ and mass energy density $\hmass$ space and time sections (with ground contribution removed) for elastic meson-meson scatterings with $\mass=0.6$ and various couplings $\cpl$ (see the color bar in the last row).
        Panels \sref{fig:S(x)}--\sref{fig:pdf(x)} plots are obtained at a fixed ($\cpl$ dependent) time $t=\tf_{\cpl}$, corresponding to outgoing wave packet peaks separated by $\Delta x \approx 100$.
        Panels \sref{fig:S(x)}, \sref{fig:Sint(x)} reproduce the final time $\tf_{\cpl}$ entanglement profile, with the free contribution ($\cpl = 0$) subtracted in \sref{fig:Sint(x)};
        while \sref{fig:pdf(x)} shows the final time mass energy density.
        Panels \sref{fig:S(t)}, \sref{fig:Swps(t)} show the midchain entanglement as a function of time; \sref{fig:Swps(t)} in an enlargement of its decaying tail, approaching the extrapolated asymptotic value,
        \begin{math}
            \ent(\infty, L/2)=\Sint+\Swps(\infty) \approx \Sint
        \end{math}.
    }
\end{figure}
The ground state and the wave packets completely explain the entanglement observed in our simulations of the free kinematics.
In particular,
if the system is cut outside the support of the outgoing wave packets, the entanglement entropy of the bipartition comes from the ground state only.
When the interaction is switched on ($\cpl > 0$) this is no longer true, because additional entanglement is generated by the dynamics.
In the elastic scattering regime explored with our simulations, the final time entanglement entropy profiles are characterized by a uniform plateau in the region enclosed between the two outgoing wave packets; we thus interpret the dynamically generated entanglement as entanglement between the scattering products.
The entanglement plateau is clearly visible in \cref{fig:S(x)} and especially \cref{fig:Sint(x)}, where we subtract the contribution from the $\cpl=0$ simulation.
For $\cpl>0$, entanglement is present also in the middle of the chain, where the mass energy density, shown in \cref{fig:pdf(x)}, vanishes.

\begin{figure}
    \begin{subcaptions}{\includegraphics{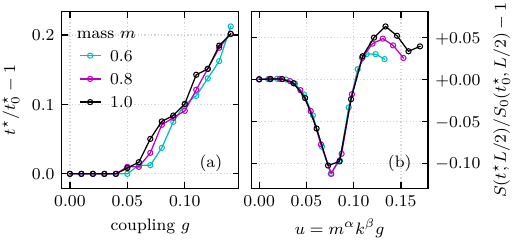}}
        \oversubcaption{0.0, 0.0}{\label{fig:Smax(g)_arg}}
        \oversubcaption{0.5, 0.0}{\label{fig:Smax(g)_val}}
    \end{subcaptions}
    \caption{\label{fig:Smax(g)}%
        Time $t^{\star}$ \sref{fig:Smax(g)_arg} and entanglement $S(t^{\mathrlap{\star}},\cells/2)$ \sref{fig:Smax(g)_val} corresponding to the peak of the midchain entanglement profiles, for different masses $\mass$ and couplings $\cpl$.
        The ordinate represents the relative deviation from the values obtained for the free ($\cpl=0$) simulation, namely $t^{\mathrlap{\star}}_0$ and $S_0(t^{\mathrlap{\star}}_0,\cells/2)$.
    }
\end{figure}
\begin{figure}
    \begin{subcaptions}{\includegraphics{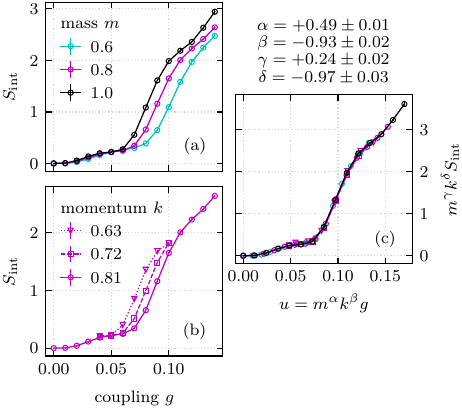}}
        \oversubcaption{0.0, 0.1}{\label{fig:Sint_gm}}
        \oversubcaption{0.0, 0.5}{\label{fig:Sint_gk}}
        \oversubcaption{0.5, 0.33}{\label{fig:Sint_gmk_universal}}
    \end{subcaptions}\hspace{0.8em}
    \caption{\label{fig:Sint(g)}%
        Growth of the interaction entanglement $\Sint$ with the coupling $\cpl$, for different masses $\mass$ \sref{fig:Sint_gm} and mean momenta $\pm\kSint$ of the initial mesons \sref{fig:Sint_gk}.
        Panel \sref{fig:Sint_gmk_universal} collects all data points from \sref{fig:Sint_gm}--\sref{fig:Sint_gk} and shows the universal behavior captured by \cref{eq:Sint_univ}.
        Error bars represent a $\pm\delta\Sint$ uncertainty, $\delta\Sint$ being the difference between the extrapolated $\Sint$ and the final midchain entanglement (with ground contribution removed).
    }
\end{figure}
The time profile of the midchain entanglement in \cref{fig:S(t)} shows that the correlation between the outgoing mesons is produced by the interactions, during the collision.
Indeed, in the initial stage of the evolution, when the
incoming mesons approach one another, no entanglement is present between them.
However, once their wave packets overlap, the associated entanglement is detected at the middle of the chain.
This process gets delayed as the coupling is increased (see \cref{fig:Smax(g)_arg}), likely due to a combination of factors, such as a deflection of the meson trajectories and the takeover of the interaction-generated entanglement.
In \cref{fig:Smax(g)_val} we plot the peak midchain entanglement $\ent(t^{\mathrlap{\star}},\cells/2)$ as a function of $\univarg=\mass^\alpha \kSint^\beta \cpl$, where $\kSint=0.81$ is the absolute value of the mean momentum of the initial mesons and the exponents $\alpha$, $\beta$ are reported in \cref{fig:Sint(g)}.
$\ent(t^{\mathrlap{\star}},\cells/2)$ is decreasing at small $\univarg$, but above the threshold value
\begin{math}
    \univstar \approx 0.08
\end{math}
it rapidly increases again.
Nevertheless, it is only after the collision that the fundamental distinction between the free and interacting cases\emdash{}namely, the entanglement of the scattering products\emdash{}emerges.

\Cref{fig:Sint(g)} shows the the coupling dependence of the entanglement between the outgoing mesons generated by the interaction, $\Sint$, for different values of the bare fermion mass $\mass$ and of the mean momenta $\pm\kSint$ of the initial mesons.
To quantify $\Sint$ we assume that all the entanglement due to the interaction is produced during the collision and remains approximately constant afterwards.
Thus, in the final stage of the evolution, the midchain entanglement can be decomposed as a constant contribution $\Sgrn+\Sint$, due to the ground state and the interaction, plus a decaying component $\Swps(t)$, attributed to the tails of the outgoing meson wave packets.
We thus evolve the system until the mesons are spatially separated by $\Delta x \gtrsim 100$ and fit the midchain entanglement values sampled in the final $\Delta t = 40$ time interval using the expression provided by \cref{eq:wavepacket_entanglement_model} for $\xx=\cells/2$, plus a constant background.
Since we always remove the ground state contribution, the value of the background is precisely the extrapolated $\Sint$.
The fits are reproduced in \cref{fig:Swps(t)}.
According to \cref{fig:pdf(x)}, for the highest $\cpl$ values the outgoing wave packets are not Gaussian, as assumed in \cref{eq:wavepacket_entanglement_model}, thus the fits are only partially justified.
Yet, especially for stronger couplings, towards the end of the evolution the wave packets give a small relative contribution to the midchain entanglement and its final value is already an accurate estimate of $\Sint$.

We find that the entanglement produced by the interaction\emdash{}i.e. the amount of of quantum correlations between the outgoing mesons\emdash{}increases with the coupling, as expected.
Moreover, we find that the collapse of the interaction entanglement curves on a unique curve $F(\univarg)$ shown in \cref{fig:Sint_gmk_universal} is described by
\begin{equation}\label{eq:Sint_univ}
    \Sint(\cpl, \mass, \kSint) = \mass^{-\gamma} \kSint^{-\delta} F(\univarg)
    \ , \quad
    \univarg = \mass^\alpha \kSint^\beta \cpl
    \ .
\end{equation}
The optimal exponents $\alpha,\beta,\gamma,\delta$ in the least-squares sense are reported in \cref{fig:Sint_gmk_universal}.
They are obtained minimizing the residual sum of squares of the rescaled data points from a common interpolating polynomial of degree 10.
\Cref{eq:Sint_univ} allows us to express our findings in terms of $\univarg$ only: after an initial gentle growth for small $\univarg$, at $\univstar$ we observe an abrupt increase in the slope of $\Sint$, which then stays constant up to
\begin{math}
    \univarg \approx 0.12
\end{math},
at which point the slope diminishes again.

\section{Conclusions and outlook}
We have shown that Tensor Network (TN) methods are viable numerical techniques for the simulation of the dynamics of $(1+1)$-dimensional lattice Quantum Electrodynamics (\qed{}).
We identified a Matrix Product State (MPS) representation of the stable particle states of the theory, namely meson bound states.
We performed real-time TN simulations of the elastic collision of a pair of mesons for moderately week couplings ($\cpl/\mass \lesssim  1/4$) and large lattice sizes ($\spc*\cells\gg\corrlen$, $\corrlen$ being the correlation length of the model).
In the simulated processes the entanglement growth is perfectly sustainable, attesting that classical TN algorithms are capable of attacking this problem efficiently and accurately.

In \cref{sec:initial_state,app:MPOs} we elaborated a protocol for the preparation of an MPS consisting of some approximate meson wave packets.
The protocol is based on the exact solution of the theory of free staggered fermions, given in \cref{app:free_sol}.
With this regard, let us stress that rather than the breakdown of the numerical tools employed, it is only our approximation of the meson states that prevented us from studying the \qed{} dynamics at stronger couplings.
Recently developed numerical techniques, such as tangent space MPS methods \cite{Vanderstraeten2019TangentSpaceMethods,Haegeman2016UnifyingTimeEvolution,Haegeman2014GeometryMatrixProduct,Haegeman2013PostMatrixProduct,Haegeman2013ElementaryExcitationsGapped,Vanderstraeten2014SMatrixMatrix} and in particular \cite{VanDamme2021RealTimeScattering}, should furnish a valid replacement for our (perturbative) initial state preparation protocol.
We expect the identification of the exact particle excitations of the model to provide access to stronger coupling regimes and thus, likely, more diverse dynamical processes.
Given the resources required by our simulations, we expect the dynamics of more complex gauge theories to be accessible via tensor network simulations as well.
Examples that are worth looking into are theories involving multiple matter field flavors and non-Abelian gauge groups \cite{Rico2014TensorNetworksLattice,Silvi2014LatticeGaugeTensor,Banuls2020SimulatingLatticeGauge,Silvi2019TensorNetworkSimulation}.

In \cref{sec:dynamics} we put forward two strategies for analyzing the momentum content of the system and applied them to identify the momenta of the collision products in our scattering simulations.
The first strategy involves the computation of a specific 2-point connected correlation function.
The second relies on the evaluation of a set of finite-time scattering amplitudes, obtained projecting the system state on a family of carefully constructed wave packets whose properties are discussed in \cref{app:heaviside_wavepacket}.
Building upon these scattering amplitudes, we gave a prescription for extracting continuum \smatrix{} elements from tensor network lattice simulations.
We believe that its implementation is within the reach of the currently available numerical tools, with reasonable computational resources.
For an alternative approach, based on the Lippmann-Schwinger formalism, see \cite{Vanderstraeten2015ScatteringParticlesQuantum}.

In \cref{sec:entanglement} we analyzed the evolution of the entanglement content of the system during our simulations of elastic meson-meson collisions.
We found that the entanglement entropy observed in these scattering processes can be decomposed in three, approximately additive, distinct contributions, namely the following:
\begin{enumerate}
    \item Vacuum entanglement: a uniform layer of entanglement produced by the correlations in the ground state of the theory.
    \item Intraparticle entanglement: entanglement bumps in the regions where particle wave packets are localized.
    \item Interparticle entanglement: a dynamically generated entanglement string correlating the products of the collision.
\end{enumerate}
After having derived approximate analytical expressions that model a particle internal entanglement, we focused on the inter particle entanglement $\Sint$.
We studied its growth with the interaction strength $\cpl$ for different values of the bare mass parameter $\mass$ and mean momentum $\kSint$ of the initial meson wave packets.
In the explored parameters region, we found a phenomenological relation describing the interplay of the mass, the coupling, the meson momenta, and the interaction entanglement $\Sint$.
Precisely, by \cref{eq:Sint_univ},
\begin{equation}
    \Sint = \mass^{-0.24} \kSint^{0.93} F(u)
    \ ,
    \quad
    \univarg = \mass^{0.49} \kSint^{-0.97} \cpl
    \ ,
\end{equation}
where $F$ is the function plotted in \cref{fig:Sint_gmk_universal}.
At fixed $\mass$ and $\kSint$, an initial slow growth of $\Sint$ for small values of the coupling $\cpl$ is followed by a rapid and steady rise, for $\univarg > \univstar \approx 0.08$.
The above relation offers a testbed for the investigation of the relation between entanglement and scattering amplitudes, motivating additional real-time dynamical investigations of scattering events.

\begin{acknowledgments}
    We gratefully acknowledge insightful discussions with S. Pascazio and P. Silvi.
    Calculations were performed using the TeNPy Library \cite{Hauschild2018EfficientNumericalSimulations}; CloudVeneto is acknowledged for the use of computing facilities.
    This work is partially supported by MIUR (through PRIN 2017) and fondazione CARIPARO, the INFN project QUANTUM, the European Union's Horizon 2020 research and innovation programme under Grant Agreements No. 817482 (PASQuanS) and No. 731473 (QuantERA, through the  QTFLAG and QuantHEP projects), and the DFG project TWITTER.
\end{acknowledgments}

\appendix

\section{Exact solution of staggered fermions}\label{app:free_sol}
In this Section we solve exactly the lattice theory of free staggered fermions, defined by \cref{eq:Hfree_x}, for periodic boundary conditions.
An analogous derivation carries over in the thermodynamic limit, with momentum sums replaced by integrals over the Brillouin zone.
We compute the energy-momentum spectrum, construct the Fock space, and finally study some features of the infinite mass limit, $\mass\to\infty$, that are relevant for this work.

Let us first introduce some notation.
We define the (complementary) orthogonal projectors
\begin{subequations}
    \begin{equation}
        \projxeo = \frac{1+(-1)^{\spcfrac*{\xx}+\eo}}{2}
        \,,\quad
        \eo\in\Zn[2]
        \,.
    \end{equation}
    Via pointwise product, $\projxeo[\xx0]$ ($\projxeo[\xx1]$) projects a field on its even (odd) sublattice component.
    On top of the usual
    \begin{align}
        \projxeo[\xx\eo[2]]\projxeo[\xx\eo[3]] = \projxeo[\xx\eo[2]] \kroneo
        \,,\quad
        \sum\nolimits_{\eo}\projxeo = 1
        \,,
    \end{align}
    we have the following useful properties:
    \begin{align}
        \stgsgn\projxeo = (-1)^{\eo}\projxeo
        \,,\quad
        \projxeo[(\xx+\spc)\eo] = \projxeo[\xx(\eo+1)]
        \,.
    \end{align}
\end{subequations}
With these projectors we introduce the staggered doublet field
\begin{math}
    \dblx = (\dblxeo)^\phdag_{\eo\in\Zn[2]}
\end{math},
\begin{math}
    \dblxeo = \projxeo\stgx
\end{math},
\begin{math}
    \stgx = \sum_{\eo}\dblxeo
\end{math};
and its Fourier transform $\dblk$,
\begin{subequations}
    \begin{align}\label{eq:dbl_k}
        \dblk & = \fouriercoef\,\sumx e^{-i\kk\xx} \dblx
        \,,                                              \\
        \dblx & = \fouriercoef\,\sumk e^{+i\kk\xx} \dblk
        \,.
    \end{align}
\end{subequations}
In momentum space $\projxeo$ becomes
\begin{equation}
    \projxeo[\kk\eo]
    = \sqrt{2\pi} \, \kspc^{-1} \frac{\kronk[\kk0]* + (-1)^{\eo}\kronk[\kk\kperiod]*}{2}
    \,,
\end{equation}
projecting on the $\kperiod$-periodic and $\kperiod$-antiperiodic parts of a field through convolution:
\begin{equation}\label{eq:dblk_periodicity}
    \dblkeo = \fouriercoef \projxeo[\kk\eo]\ast\stgk
    \,,\quad
    \dblkeo[(\kk+\kperiod*)\eo] = (-1)^{\eo}\dblkeo
    \,.
\end{equation}
Often $\kperiod$ periodicities make it convenient to work with momenta contained in
\begin{math}
    \stglatk = \latk\cap [-\pi/2\spc*, \pi/2\spc*[
\end{math};
we denote sums restricted to $\stglatk$ by $\stgsumk*[]$.

\subsection{Spectrum}
In terms of the doublet $\dblx$, the Hamiltonian reads
\begin{subequations}\label{eq:Hfree_k_eig}
    \setlength\arraycolsep{0pt}%
    \begin{align}
        H & =\frac{\spc*}{2} \sumx* \left[
            \dblx[x+\spc]* \hspace{-1pt}\begin{pmatrix} 0 & \spcfrac*{i} \\ \spcfrac*{i} & 0 \end{pmatrix}\hspace{-1pt} \dblx +
            \dblx* \hspace{-1pt}\begin{pmatrix} +\mass & 0 \\ 0 & -\mass \end{pmatrix}\hspace{-1pt} \dblx + \hc
            \right]
        \nonumber
        \intertext{and Fourier transforming,}
        H & =\frac{1}{2} \sumk \left[ \dblk* \begin{pmatrix}
                +\mass                    & i\spc*^{-1}e^{-i\kk\spc*} \\
                i\spc*^{-1}e^{-i\kk\spc*} & -\mass
            \end{pmatrix} \dblk + \hc \right] \nonumber \\
          & = \sumk  \dblk* \begin{pmatrix}
            +\mass & \kart  \\
            \kart  & -\mass
        \end{pmatrix} \dblk \label{eq:Hfree_k_quadraticforms}
        \,.
    \end{align}
    Otherwise stated, the momentum space Hamiltonian is a sum of quadratic forms, one for each Fourier mode $\kk$,
    diagonalized by the unitary boost matrices $\boost$,
    \begin{equation}
        \boost = \frac{\sqrt{\smpp}}{\sqrt{2\pulse}}
        \begin{pmatrix}
            1    & +\vk \\
            -\vk & 1
        \end{pmatrix}
        ,\quad
        \boost \boost* = 1
        \,;
    \end{equation}
    with,
    \begin{equation}\label{eq:dispersion}
        \pulse = \sqrt{\mass^2 + \kart^2}
        ,\quad
        \vk = \frac{\kart}{\smpp}
        \,.
    \end{equation}
    Notice that $\vk\to0$ and $\boost\to\mathbf{1}$ for $\mass\to\infty$.
    Introducing
    \begin{equation}
        ( \posk , \negk[-k]* ) = \sqrt{2}\, \boost \dblk
    \end{equation}
    the Hamiltonian becomes
    \begin{equation}
        H = \stgsumk \pulse \big[ \posk*\posk - \negk\negk* \big] \label{eq:Hfree_k_divergent}
        \,.
    \end{equation}
\end{subequations}
In the last step we recognized that the quadratic forms of the $\kk$ and $\kk+\kperiod*$ modes in \cref{eq:Hfree_k_quadraticforms,eq:Hfree_k_divergent} are degenerate and restricted the sum to $\stglatk$.

Explicit expressions for $\posk*$ and $\negk*$ in terms of the position space fields, $\stgx*$ and $\stgx$, are obtained combining the previous results.
We report them below for reference:
\begin{subequations}\label{eq:ab_def}
    \begin{align}
        \label{eq:a_def}
        \posk* & = \frac{\sqrt{2}}{\sqrt{2\pi}}\frac{\sqrt{\smpp}}{\sqrt{2\pulse}} \,
        \sumx e^{i\kk\xx} \big[\projxeo[\xx0]+\vk\projxeo[\xx1]\big] \stgx* \,,       \\
        \label{eq:b_def}
        \negk* & = \frac{\sqrt{2}}{\sqrt{2\pi}}\frac{\sqrt{\smpp}}{\sqrt{2\pulse}} \,
        \sumx e^{i\kk\xx} \big[\projxeo[\xx1]-\vk\projxeo[\xx0]\big] \stgx \,.
    \end{align}
\end{subequations}
Moreover, by \cref{eq:Hfree_k_eig,eq:dblk_periodicity},
\begin{equation}
    \posk[\kk+\kperiod*]*=+\posk*
    \,,\qquad
    \negk[\kk+\kperiod*]*=-\negk*
    \,.
\end{equation}
To avoid redundancies we take the operators $\posk$ and $\negk$ as defined only for $\kk\in\stglatk$.
With this caveat, they satisfy the same canonical anticommutation relations of $\stgx$,
\begin{equation}\label{eq:ab_acomm}
    \acomm{\posk[2]}{\posk[3]*} = \kronk
    \,,\quad
    \acomm{\negk[2]}{\negk[3]*} = \kronk
    \,,
\end{equation}
while other fundamental commutators vanish.
Indeed,
\begin{align}
    \acomm*{\dblkeo[2]}{\dblkeo[3]*} \!
     & = \frac{a^2}{2\pi}\sumx*[2;3] e^{-i\kk[2]\xx[2] +i\kk[3]\xx[3]} \acomm*{\stgx[2]}{\stgx[3]*} \! \projxeo[2]\projxeo[3] \nonumber \\
     & = \frac{a^2}{2\pi}\sumx*[2;3] e^{-i(\kk[2]-\kk[3])\xx[2]} \kronx \kroneo \projxeo[2] \nonumber                                   \\
     & = \kspc^{-1} \kroneo \frac{\kronk* + (-1)^{\eo[2]}\kronk[\kk[2](\kk[3]+\spcfrac*{\pi})]*}{2} \label{eq:dblk_acomm}
    \,.
\end{align}
On $\stglatk$ only the first Kronecker delta in \cref{eq:dblk_acomm} contributes and \cref{eq:ab_acomm} follows from the unitarity of $\boost$.

Recalling the $\negk, \negk*$ anticommutator in \cref{eq:ab_acomm}, \cref{eq:Hfree_k_divergent} can be rewritten as a positive semidefinite operator minus a constant.
The constant term, namely $\stgsumk*\pulse$, incorporates both ultraviolet and infrared divergencies when the respective regulators are released, making the continuum limit ill defined.
Dropping it by a normal ordering prescription, the free staggered fermions energy-momentum operator $P^\mu=(H, P)$ reads
\begin{equation}\label{eq:Hfree_k}
    P^\mu = \stgsumk \kk^\mu \big[ \posk*\posk + \negk*\negk \big]
    \,,\quad
    k^\mu = (\pulse, \kk)
    \,.
\end{equation}
\Cref{eq:Hfree_k} shows that $\posk*$ and $\negk*$ create excitations of energy-momentum $\kk^\mu$:
\begin{equation}
    \comm*{P^\mu}{\posk*} = \kk^\mu \posk*
    \,,\quad
    \comm*{P^\mu}{\negk*} = \kk^\mu \negk*
    \,.
\end{equation}
It also identifies $\pulse$ in \cref{eq:dispersion} as the staggered fermion dispersion relation.
Its derivative $\pulse'$ gives the group velocity of fermion wave packets on the lattice.

\subsection{Fock space}
The ground state of the theory, $\ket{\freevac}$, is the vacuum $P_\mu\ket{\freevac}=0$, i.e., the state with no excitations to destroy:
\begin{equation}\label{eq:fock_vacuum}
    \posk\ket{\freevac}=\negk\ket{\freevac}=0
    \,,\quad
    \braket{\freevac} = 1
    \,.
\end{equation}
The Fock space of all the (normalized and antisymmetrized) single and multiparticle states of the theory is generated by acting on $\ket{\freevac}$ with products of creation operators $\posk*$ and $\negk*$:
\begin{equation}\label{eq:fock_states}
    \ket{\kk[3]_{N}\ldots\kk[3]_{1};\kk[2]_{M}\ldots\kk[2]_{1}} =
    \negk[\kk[3]_N]* \!\! \cdots \negk[\kk[3]_1]* \,
    \posk[\kk[2]_M]* \!\! \cdots \posk[\kk[2]_1]*
    \ket{\freevac}
    \,,
\end{equation}
with $\kk[2]_{i},\kk[3]_{j}\in\stglatk$.
For instance, by \cref{eq:ab_acomm,eq:fock_vacuum},
\begin{equation}\label{eq:fock_orthonormal}
    \braket{\kk[2]}{\kk[2]'} = \kronk[\kk[2;2]']
    \,,\quad
    \braket{\kk[3]}{\kk[3]'} = \kronk[\kk[3;3]']
    \,.
\end{equation}
In terms of $\posk$ and $\negk$, the global $\Uone$ charge $Q$ reads
\begin{equation}\label{eq:Qfree_k}
    Q
    = \sumx \dblx*\dblx
    = \sumk \dblk*\dblk
    = \stgsumk \big[ \posk*\posk - \negk*\negk \big]
    \,,
\end{equation}
where in the last step we have normal ordered, discarding the half-filling constant ${\cells}/{2}$.
\Cref{eq:Qfree_k} shows that $\negk[]$-type excitations (antifermions) are the antiparticles of $\posk[]$-type excitations (fermions).

\subsection{Infinite mass limit}
The theory of free staggered fermions simplifies greatly when $\mass\gg\spcfrac*{1}$.
For instance, recalling the derivation in \cref{eq:Hfree_k_eig}, in the $\mass\to\infty$ limit \cref{eq:ab_def} reduces to
\begin{subequations}\label{eq:ab_def_limit}
    \begin{align}
        \posk* & = \frac{\sqrt{2}}{\sqrt{2\pi}} \:\sumx e^{i\kk\xx} \projxeo[\xx0] \stgx* \,, \\
        \negk* & = \frac{\sqrt{2}}{\sqrt{2\pi}} \:\sumx e^{i\kk\xx} \projxeo[\xx1] \stgx \,.
    \end{align}
\end{subequations}
We conclude that fermions (antifermions) excitations are supported on the even ($\sys{E}$) and odd ($\sys{O}$) sublattices.
Moreover, inserting \cref{eq:ab_def_limit} in the expressions of $\poswp*$ and $\negwp*$ from \cref{eq:free_wavepacket_creation} yields
\begin{equation}\label{eq:wpx_limit}
    \wpx^{\poswpSym(\negwpSym)} = \projxeo[\xx,0(1)] \frac{\sqrt{2}}{\sqrt{2\pi}}\stgsumk e^{i\kk\xx} \wpk
    \,.
\end{equation}
Therefore $\wpx^{\poswpSym}$ ($\wpx^{\negwpSym}$) vanishes on $\sys{O}$ ($\sys{E}$), while its restriction to $\sys{E}$ ($\sys{O}$) is the inverse Fourier transform of $\wpk$.

Consequently, recalling \cref{eq:wavepacket_normalization},
$\sabs*{\wpx^{\poswpSym,\negwpSym}}$ become PDFs on $\latx$ for $\mass\to\infty$.
Their CDFs read
\begin{equation}\label{eq:free_cdf_lim}
    \cdf^{\poswpSym(\negwpSym)}
    = \sumx[\xx[2]<\xx] \sabs*{\wpx[2]^{\poswpSym(\negwpSym)}}
    = 2\sumx[\mathclap{\substack{\xx[2]<\xx\\\xx[2]\in\sys{E(O)}}}] \:\sabs{\fouriercoef\smash{\stgsumk} e^{i\kk\xx[2]} \wpk}
    ;
\end{equation}
whose thermodynamic and continuum limit is
\begin{equation}\label{eq:free_cdf_limlim}
    \cdf =
    \int_{\mathrlap{-\infty}}^{\mathrlap{\xx}}\dd{\xx[2]}
    \sabs{\fouriercoef\int_{\mathrlap{-\infty}}^{\mathrlap{+\infty}}\dd{\kk} e^{i\kk\xx[2]} \wpk}\!
    =
    \int_{\mathrlap{-\infty}}^{\mathrlap{\xx}}\dd{\xx[2]}
    \pdf[2]
\end{equation}
for both fermions and antifermions.
Note that $\wpk[\xx[2]]$ in \cref{eq:free_cdf_limlim} is the inverse Fourier transform of $\wpk$.
These results make the infinite mass limit convenient for the characterization of the real space entanglement entropy of particle wave packets, carried out in \cref{sec:entanglement}.

In \cref{sec:entanglement}
we also study the time evolution of $\cdf$ in \cref{eq:free_cdf_limlim}.
For a fermion or antifermion wave packet,
\begin{math}
    e^{-iHt} \sumk*\wpk\ket{\kk} = \sumk*e^{-i\pulse t} \wpk\ket{\kk}
\end{math}.
Expanding the dispersion relation of \cref{eq:dispersion} in $1/\spc*\mass$ yields
\begin{equation}
    \pulse = \spc*^{-1}\left[ \spc*\mass + \frac{\sin^2(\spc*\kk)}{2\spc*\mass} + \order*{(\spc*\mass)^{-3}} \right]
    \,.
\end{equation}
Disregarding the inconsequential global phase $e^{-i\mass t}$, the proper wave packet evolution comes from subleading terms in the expansion and takes place on timescales $\spcfrac*{t}\sim\spc*\mass$.
Accordingly, sending $\mass\to\infty$ we implicitly assume $t=\tau\spc*\mass$ with $\tau$ finite.

\begin{figure*}[t]
    \includegraphics{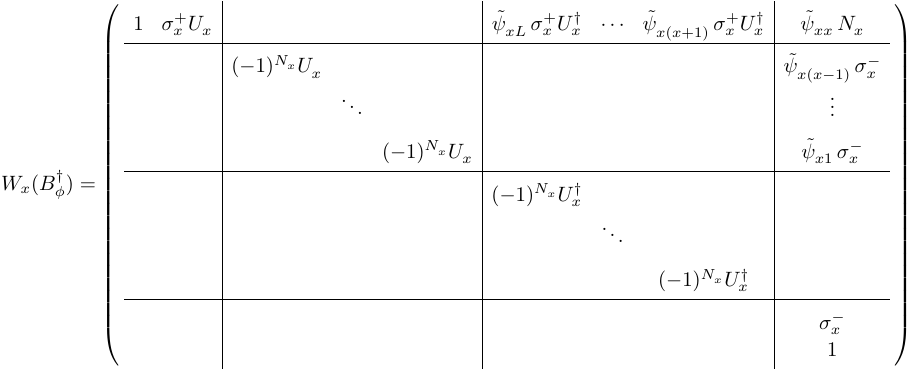}
    \caption{\label{fig:meson_MPO}%
        $W_x$ matrices of the meson wave packet creation MPO.
        Empty entries represent null operators.
    }
\end{figure*}

\section{Wave packet creation MPOs}\label{app:MPOs}

We write the MPO \cite{Pirvu2010MatrixProductOperator} representation of a generic many-body operator $O$ as
\begin{equation}\label{eq:mpo_ansatz}
    O = w_0 W_1 W_2 \cdots W_{\cells} w_{\cells}
    \,.
\end{equation}
Here $W_{\xx}=W_{\xx}(O)$ are matrices whose entries are operators acting nontrivially only on the local Hilbert space of site $\xx$, while the vectors $w_0$ and $w_{\cells}$ are introduced to obtain a uniform bulk and read
    {
        \delimiterfactor=800
        \begin{equation*}
            w_0=\begin{pmatrix}
                1 & 0 & 0 & 0
            \end{pmatrix}
            ,\qquad
            w_\cells=\begin{pmatrix}
                0 & 0 & 0 & 1
            \end{pmatrix}^T
            .
        \end{equation*}
    }
It is a known result that short range interactions, such as those appearing in the Hamiltonians of \cref{sec:models}, can be represented exactly as MPOs with small bond dimension \cite{Hauschild2018EfficientNumericalSimulations}.
In this Section we provide an explicit MPO representation of the fermion, antifermion and meson wave packet creation operators in \cref{eq:fermion_wavepacket_creation,eq:meson_wavepacket_creation}.

\subsection{Fermion and antifermion MPOs}

In terms of the Jordan-Wigner matrix representation of the staggered fermion operators, the wave packet creation operator $\poswp$ of \cref{eq:fermion_wavepacket_creation} reads
\begin{equation}\label{eq:fermion_wavepacket_creation_explicit}
    \poswp* = \sumx \wpx^{\poswpSym} \Big[\prod_{y<x} (-1)^{\occ[y]}\Big] \sigma^+_x
    \,.
\end{equation}
The nonvanishing matrix elements of the single-site operators involved are
\begin{math}
    \mel{1}{\sigma^{+}}{0}
    =
    \mel{1}{\occSym}{1} = 1
\end{math}.
Kronecker products of local operators with identities acting on other local Hilbert spaces are left implicit.

Despite the nonlocal Jordan-Wigner strings, the $\poswp*$ operator in \cref{eq:fermion_wavepacket_creation_explicit} admits a simple MPO representation.
In the notations of \cref{eq:mpo_ansatz}, its $W_{\xx}$ matrices read
\begin{equation}\label{eq:res_free_mpo_creation}
    W_x({\poswp*}) =
    \begin{pmatrix}
        (-1)^{\occ} & \wpx^{\poswpSym}\, \sigma^+_x \\
        0           & 1                             \\
    \end{pmatrix}
    \,.
\end{equation}
The analogous result for the antifermion operator $\negwp*$ is obtained via the substitution $\sigma^+\to\sigma^-=(\sigma^+)^\dagger$.

\subsection{Meson MPO}
After the Jordan-Wigner transformation and some algebraic manipulations
the meson wave packet creation operator in \cref{eq:meson_wavepacket_creation} reads
\begin{equation*}\label{eq:meson_wavepacekt_creation_explicit}
    \meswp* =
    \spc*^2\sumx*[2;3] \wpxx[2;3] \begin{dcases}
        \sigma^+_{\xx[2]} \cmp[2]*
        \left.\Big[
            \prod_{\xx=\xx[2]}^{\xx[3]} (-1)^{\occ} \cmp*
            \Big]\right.
        \sigma^-_{\xx[3]}
        \,, & \xx[2]<\xx[3] \,;
        \\
        \occ[2]
        \,, & \xx[2]=\xx[3] \,;
        \\
        \sigma^-_{\xx[3]} \cmp[3]
        \left.\Big[
            \prod_{\xx=\xx[3]}^{\xx[2]} (-1)^{\occ} \cmp
            \Big]\right.
        \sigma^+_{\xx[2]}
        \,, & \xx[2]>\xx[3] \,;
    \end{dcases}
\end{equation*}
where we set $\cmp=\cmpR$ for compactness of notation.

The $W_x$ matrices of the MPO representation of this operator are shown in \cref{fig:meson_MPO}.
These have linear dimension $\cells+2$, a fact that might make the contraction with an MPS quite resource heavy for long chains.
However, the MPO can be compressed, numerically \cite{Schollwoeck2011DensityMatrixRenormalization} or analytically, by discarding the rows and columns related to irrelevant amplitudes, $\wpxx[2;3]\ll 1$.
As an example, in preparing the mesons depicted in \cref{fig:wavepacket} we truncate each meson amplitude $\wpxx$ outside the dashed ellipses of \cref{fig:wavepacket}.

{
\section{Half-chain plane waves}\label{app:heaviside_wavepacket}
\def\ieps{\lim_{\ \epsilon\to 0^{+}}}%
\def\iepsint{\ieps \int_{\mathrlap{-\infty}}^{\mathrlap{+\infty}}}%
\def\xmean{y}%
\def\kmean{q}%
\def\kwp{\theta^{\kmean}}%
\def\kwpL{\theta^{\kmean}}%
\def\xwp{\phi^{\xmean}}%
Consider a continuum theory in one space dimension with single particle energy-momentum eigenstates $\ket*{k}$, $\braket*{k'}{k}=\delta(k'-k)$.
For simplicity we assume the theory has only one particle specie.
The momentum space amplitude of a wave packet
\begin{math}
    \ket*{\kwp} = \int\!\dd{k}\kwp(k)\ket{k}
\end{math},
completely delocalized in the $x>0$ space region and peaked at momentum $\kmean$, reads
\begin{equation}
    \begin{aligned}
        \kwp(k) & = \frac{1}{2\pi} \iepsint\dd{x}\,e^{-ikx}\,e^{i(\kmean + i\epsilon)x}\,\Theta(x)   \\
                & = \frac{1}{2\pi i} \ieps \frac{1}{k - \kmean - i\epsilon}                          \\
                & = \frac{1}{2} \delta(k - \kmean) + \frac{1}{2\pi i}\mathcal{P}\frac{1}{k - \kmean}
        \,.
    \end{aligned}
\end{equation}
Here $\Theta(x)$ is the Heaviside step function and $\mathcal{P}$ denotes the Cauchy principal value, while an $i\epsilon$ prescription has been introduced for formal convergence.
This $\kwp(k)$ should be interpreted in the sense of distributions, it is unnormalizable and the variance of $\sabs{\kwp(k)}$ is undefined.
Nonetheless, as we now show, $\bra*{\kwp}$ projects (sufficiently well behaved) wave packets peaked at $x>0$ in position space, on their component, i.e., wave packet amplitude, of momentum $\kmean$.

Let
\begin{math}
    \ket*{\xwp} = \int\dd{k} e^{-ik\xmean} \phi(k) \ket{k}
\end{math}
be a wave packet peaked at $x=y$ such that, as a complex function, $\phi(k)$ has no singularities and \begin{math}
    \abs{\phi(k)} \to 0
\end{math}
for $\abs{k}\to\infty$
(these criteria are satisfied, e.g., by a complex Gaussian $e^{-\sabs{k}}$).
We want to compute
\begin{equation}
    \braket*{\kwp}{\xwp} = -\frac{1}{2\pi i}\iepsint\dd{k} \frac{e^{-ik\xmean}\phi(k)}{k - \kmean + i\epsilon}
    \,.
\end{equation}
For large enough $\abs{k}$,
\begin{equation}
    \abs{\frac{\phi(k)}{k - \kmean + i\epsilon}} \leq \frac{\abs{\phi(k)}}{\abs{k} - \abs{\kmean - i\epsilon}} < \abs{\phi(k)}
    \,;
\end{equation}
therefore, we can invoke Jordan's lemma closing the contour of integration in the lower half (upper half) of the complex plane for $y>0$ ($y<0$).
The integrand has a single pole at $k = \kmean - i\epsilon$ with
\begin{equation}
    \ieps\;\underset{k=\kmean-i\epsilon}{\Res\limits}\;\frac{e^{-ik\xmean}\phi(k)}{k - \kmean + i\epsilon}
    = e^{-i\kmean\xmean}\phi(\kmean)
\end{equation}
and the residue theorem yields
\begin{equation}
    \braket*{\kwp}{\xwp}
    = \Theta(y) e^{-i\kmean\xmean}\phi(\kmean)
    \,,
\end{equation}
which is exactly the anticipated claim.
The analogous result for wave packets delocalized in the $x<0$ region follows by parity symmetry.
}

\end{document}